\documentclass{emulateapj}

\usepackage{graphicx,amssymb,verbatim,times}

\newcommand\hii{\hbox{H$\rm\small II$~}}

\newcommand\nii{\hbox{[N{\small$\rm II$}]}}
\newcommand\oiii{\hbox{[O{\small$\rm III$}]}}
\newcommand\sii{\hbox{[S{\small$\rm II$}]}}
\newcommand\oi{\hbox{[O{\small$\rm I$}]}}

\newcommand\niib{\hbox{[N{\scriptsize$\rm II$}]}}
\newcommand\oiiib{\hbox{[O{\scriptsize$\rm III$}]}}
\newcommand\siib{\hbox{[S{\scriptsize$\rm II$}]}}
\newcommand\oib{\hbox{[O{\scriptsize$\rm I$}]}}

\newcommand\ha{H$\alpha$}
\newcommand\hb{H$\beta$}

\slugcomment{}

\shorttitle{Local Analogs to Lyman Break Galaxies}
\shortauthors{Overzier, R.A. et al.} 
\begin{document}


\title{Local Lyman break galaxy analogs:\\
The impact of massive star-forming clumps on the interstellar medium and
the global structure of young, forming galaxies}


\author{Roderik A. Overzier\altaffilmark{1}, 
Timothy M. Heckman\altaffilmark{2}, 
Christy Tremonti\altaffilmark{3},
Lee Armus\altaffilmark{4}, 
Antara Basu-Zych\altaffilmark{5}, 
Thiago Gon\c calves\altaffilmark{6}, 
R. Michael Rich\altaffilmark{7},
D. Christopher Martin\altaffilmark{6}, 
Andy Ptak\altaffilmark{2}, 
David Schiminovich\altaffilmark{8}, 
Holland C. Ford\altaffilmark{2}, 
Barry Madore\altaffilmark{9},
Mark Seibert\altaffilmark{9}.
}
\email{overzier@mpa-garching.mpg.de}

\altaffiltext{1}{Max-Planck-Institut for Astrophysics, D-85748 Garching, Germany.}
\altaffiltext{2}{Department of Physics and Astronomy, The Johns Hopkins University, 3400 North Charles Street, Baltimore, MD 21218.}
\altaffiltext{3}{Max-Planck Institute for Astronomy, K\"onigstuhl 17, D-69117, Heidelberg, Germany.}
\altaffiltext{4}{Spitzer Science Center, California Institute of Technology, Pasadena, CA.}
\altaffiltext{5}{NASA Goddard Space Flight Center, Laboratory for X-ray Astrophysics, Greenbelt, MD 20771, USA.}
\altaffiltext{6}{California Institute of Technology, MC 405-47, 1200 East California Boulevard, Pasadena, CA 91125.}
\altaffiltext{7}{Deptartment of Physics and Astronomy, Division of Astronomy and Astrophysics, University of California, Los Angeles, CA 90095-1562, USA.}
\altaffiltext{8}{Department of Astronomy, Columbia University, MC 2457, 550 West 120th Street, New York, NY 10027.}
\altaffiltext{9}{Observatories of the Carnegie Institution of Washington, 813 Santa Barbara Street, Pasadena, California 91101, USA.}


%\altaffiltext{1}{Visiting Astronomer, Cerro Tololo Inter-American Observatory.
%CTIO is operated by AURA, Inc.\ under contract to the National Science


\begin{abstract}
  We report on the results of HST optical and UV imaging, Spitzer
  mid-IR photometry and optical spectroscopy of a sample of 30
  low-redshift ($z$$\sim$0.1 to 0.3) galaxies chosen from the SDSS and
  GALEX surveys to be accurate local analogs of the high-redshift
  Lyman Break Galaxies (LBGs). The Lyman Break Analogs (LBAs) are
  similar in stellar mass, metallicity, dust extinction, SFR, physical
  size and gas velocity dispersion, thus enabling a detailed
  investigation of many processes that are important in star forming
  galaxies at high redshift. The main optical emission line
    properties of LBAs, including evidence for outflows, are also
    similar to those typically found at high redshift. This indicates
    that the conditions in their interstellar medium are
    comparable. In the UV, LBAs are characterized by complexes of
    massive clumps of star-formation, while in the optical they most
    often show evidence for (post-)mergers and interactions. In six
  cases, we find a single extremely massive (up to several $\times
  10^9 M_{\odot}$) compact (radius $\sim 10^2$ pc) dominant central
  object (DCO). The DCOs are preferentially found in LBAs with the
  highest mid-IR luminosities ($L_{24\mu m} = 10^{10.3}$ to
  $10^{11.2}$ L$_{\odot}$) and correspondingly high star formation
  rates (15 to 100 M$_{\odot}$ per year). We show that the massive
  star-forming clumps (including the DCOs) have masses much larger
  than the nuclear super star clusters seen in normal late type
  galaxies. However, the DCOs do have masses, sizes, and densities
  similar to the excess-light/central-cusps seen in typical elliptical
  galaxies with masses similar to the LBA galaxies. We suggest that
  the DCOs form in the present-day examples of the dissipative mergers
  at high-redshift that are believed to have produced the
  central-cusps in local ellipticals (consistent with the disturbed
  optical morphologies of the LBAs). More generally, the properties of
  the LBAs are consistent with the idea that instabilities in a
  gas-rich disk lead to very massive star-forming clumps that
  eventually coalesce to form a spheroid.  Finally we comment on the
  apparent lack of energetically significant AGN in the DCOs. We
  speculate that the DCOs are too young at present to be growing a
  supermassive black hole because they are still in a
  supernova-dominated outflow phase (age less than 50
  Myr).  \end{abstract}



\section{Introduction}

One of the most remarkable features of high redshift ($z\sim2-6$)
galaxies is their great potential for vigorous star
formation. Moreover, observations with the {\it Hubble Space
  Telescope} (HST) indicate that that star formation largely occurs in
an extremely ``clumpy'' and compact mode
\citep[e.g.][]{cowie95,franx97,elmegreen05,stark08}. These clumps are
reminiscent of star-forming complexes in local \hii\ regions, but
their total masses and scale sizes are several orders of magnitudes
larger. As shown by \citet{elmegreen09}, clumps with masses of
$10^{7-9}$ $M_\odot$ and sizes of $\gtrsim$1 kpc are a very typical
constituent of extended, irregular galaxies at $z=1-4$ in the Ultra
Deep Field \citep[UDF,][]{beckwith06}. The relatively high gas velocity
dispersions observed in some high redshift galaxies
\citep[e.g.][]{law07,law09,forster09} coupled with their large
inferred gas column densities and gas fractions can explain the
formation of the clumps and their bulk properties from gravitational
instabilities in gas-rich, Jeans-unstable disks. This process may be instigated, for example, by the continuous
accretion of gas or by the delivery of gas-rich material in small
merger events \citep{bournaud09,elmegreen09}. In any case, the compactness of
the stellar clumps furthermore implies that they can form relatively
unhampered by velocity shear or supernova feedback. Simulations show
that for some galaxies the coalescence of clump systems may lie at the
root of the formation of galaxy bulges, and possibly even their
supermassive nuclear black holes if it is assumed that each clump gave
rise to its own intermediate mass black hole prior to their migration
to the center \citep{noguchi99,immeli04,elmegreen08a,elmegreen08b}.

It would be extremely valuable to study similar objects in nearby
galaxies where multi-waveband data with relatively high spatial
resolution and sensitivity would allow the relevant physical processes
to be probed in much greater detail. While local starburst galaxies used in previous studies have some similarities to 
star-forming galaxies at high redshift, there
are some important differences: local starbursts having SFRs of
$\gtrsim$100 $M_\odot$ yr$^{-1}$ are very dusty systems (the so-called
Ultra-Luminous Infrared Galaxies, ULIRGs). The intense star formation
occurs in compact, sub-kpc scale regions in the centers of interacting
galaxies, and are usually inconspicuous in the far-UV
\citep{heckman98,goldader02}. Although very dusty starbursts clearly
make up a non-negligible fraction of high redshift galaxies
\citep[e.g.][]{hughes98,huang05,papovich06,caputi07,burgarella09,daddi09},
most of the systems contributing to the cosmic SFR at high redshift
are accounted for in deep UV-selected samples of Lyman Break Galaxies (LBGs) 
\citep[e.g.][]{reddy06,reddy09,bouwens09}, indicating that the amount
of dust extinction in LBGs is typically much lower: the average ratio
of their bolometric IR-to-UV luminosities, $L_{IR}/L_{UV}$, is much
lower ($\sim10\times$) compared to local galaxies of the same
$L_{bol}$ \citep[e.g.][]{buat07,reddy06,burgarella07}.  Locally
calibrated dust corrections that are based on the reddening of the UV
continuum slope as derived by \citet{meurer99} seem to apply to
typical LBGs \citep{reddy06} with some exceptions
\citep[e.g.][]{baker01,reddy06,siana08}, and can be used to show that, 
despite their UV selection, LBGs at $z\simeq2-4$ still put out most of
their energy in the mid- and far-IR \citep{bouwens09}.

The star formation in LBGs typically occurs over the entire extent of the galaxy ranging from
a few hundred pc to several kpc
\citep[e.g.][]{franx97,giavalisco02,bouwens04,ferguson04,stark08}. The relatively low extinctions and high UV
surface brightnesses seen in LBGs are characteristic of local blue
compact dwarf galaxies (BCDs). However, typical BCDs have star
formation rates up to two orders of magnitude smaller than LBGs and occur in galaxies with much
lower mass \citep[e.g.][]{telles97,hopkins02}, although some may
resemble fainter LBGs at high redshift in their basic properties
\citep[see, e.g.,][]{ostlin09}. The main differences between
LBGs and typical local starbursts studied previously may be
due to a systematically lower metallicity and higher gas-mass fraction
in the LBGs and/or due to a different triggering mechanism
\citep[e.g.][]{erb06a,erb06b,law09}.

The ``Lyman break analogs'' (LBA) project was designed in order to
search for local starburst galaxies that share typical characteristics
of high redshift LBGs \citep{heckman05}. The LBA sample and main
properties are given in \citet{heckman05}, \citet{hoopes07},
\citet{basu-zych07,basu-zych09} and \citet[][Paper I]{overzier08}, and
we note that the LBAs are identical to the sample of ``supercompact
UV-luminous galaxies'' (ScUVLGs) referred to in the papers listed
above. In brief, the UV imaging survey performed by the Galaxy
Evolution Explorer (GALEX) was used in order to select the most
luminous ($L_{FUV}>10^{10.3}$ $L_\odot$) and most compact
($I_{FUV}>10^{9}$ $L_\odot$ kpc$^{-2}$) star-forming galaxies at
$z<0.3$. Galaxies having these high $L_{FUV}$ are already extremely
rare (local space density of $\sim10^{-5}$ Mpc$^{-3}$), and the
surface brightness requirement further reduces their absolute number
density by a factor of $\sim6$. These objects tend to be much more
luminous in the UV than typical local starburst galaxies studied
previously \citep{heckman98,meurer97,meurer99} consistent with high
SFRs and relatively little dust extinction. The median absolute
  UV magnitude of the sample is --20.3, corresponding to
  $\simeq0.5L_{z=3}^*$, where $L_{z=3}^*$ is the characteristic
  luminosity of LBGs at $z\sim3$ \citep[][i.e.,
    $M_{1700,AB}=-21.07$]{steidel99}. Analysis of their spectra from
the Sloan Digital Sky Survey (SDSS) and their spectral energy
distributions from SDSS and GALEX subsequently showed that the LBAs
are similar to LBGs in their basic global properties, including:
stellar mass, metallicity, dust extinction, SFR, physical size and gas
velocity dispersion. In \citet{overzier08} we showed that most of the
UV emission in a preliminary sample of 8 LBAs originates in highly compact burst regions in
small, clumpy galaxies that are also morphologically similar to
LBGs. We demonstrated that {\it if} LBGs at high redshift are also small merging
galaxies similar to the LBAs, unfortunately this would be very hard to
detect 
given the much poorer physical resolution and sensitivity.
Furthermore, for the present paper it is important to point out
that the LBAs also occupy an ``offset'' region in the main optical
emission line diagnostics diagram (i.e. log(\nii/\ha) versus
log(\oiii/\hb), or the ``BPT'' \citep{baldwin81} diagram) analogous to
galaxies at high redshift. This suggests that similar physical
conditions may apply to their interstellar medium (ISM) as well. In
summary, the LBAs appear indeed good local analogs to the LBGs.

In this paper we will present new results from our HST and Spitzer
imaging campaigns and spectroscopic follow-up in order to highlight
some of the remarkable properties of the LBA sample: we will compare
the UV--optical sizes and morphologies, showing evidence for massive
star-forming ``clumps'' that dominate the UV (\S3). We explore possible
connections between the starburst regions and the general properties of
the optical emission line gas (\S3.5 and \S4), as well the
relationship between the formation and evolution of massive stellar
clumps in the context of the main structural components of young
galaxies (\S5). We conclude the paper with a summary and final remarks (\S6).
Forthcoming work on the LBA sample currently in preparation includes a
detailed morphological analysis, multi-waveband modeling, mid-IR
spectroscopy, VLBI radio imaging, X-ray spectroscopy, and integral
field and long-slit optical and near-IR spectroscopy.

\section{Data}

\subsection{HST and Spitzer Imaging}

We observed 30 LBAs using HST (Programs 10920/11107, PI: Heckman). In
order to probe morphologies in the UV, twenty-four objects were
observed using the ACS Solar Blind Channel (SBC) in the filter F150LP
($\lambda_c\approx1614$\AA), 
six objects were observed using the High Resolution Camera (HRC) in
the filter F330W ($\lambda_c\approx3334$\AA).
The exposure times per object were typically $\sim$2500 s. These UV
images offer a tremendous gain in resolution over that of the GALEX
images ($\approx$$0\farcs07$ compared to $\approx$5\arcsec\ for
GALEX). Matching observations in the rest-frame optical were carried
out either using the Wide Field and Planetary Camera 2 for $\sim$3600
s in F606W ($\lambda_c\approx6001$\AA; twenty-four objects) or using
the ACS Wide Field Channel for $\sim$2200 s in F850LP
($\lambda_c\approx9170$\AA; seven objects).  The ACS and WFPC2
observations were divided into, respectively, three and six dithered
exposures and were combined using {\tt multidrizzle}
\citep{koekemoer02}.

We make use of photometric data obtained with the Multiband Imaging
Photometer (MIPS) on the {\it Spitzer Space Telescope}
at 24 $\mu$m (Program ID 20390, PI: C. Hoopes). Photometry
was obtained by performing point source extractions on the post-Basic
Calibration Data. 

\subsection{SDSS and VLT/FLAMES Spectroscopy}

We further use integrated optical emission line fluxes and flux ratios
measured from the SDSS spectra. In order to account for stellar absorption
lines, the emission-line parameters were obtained from Gaussian fits
after subtracting the continuum based on the best-fitting spectrum
using stellar population models \citep[see][]{tremonti04}.

The fraction of the total (line and continuum) light that is included
within the 3\arcsec\ diameter SDSS fiber apertures is found to be
$0.6\pm0.1$, estimated from the difference in fluxes measured in a
3\arcsec\ aperture compared to the total SDSS photometric 
aperture\footnote{The fraction of fiber-to-total-light in the SDSS
  $r^\prime$-band is estimated by $f_{fiber}=10^{-0.4(m_{r,fiber}-m_{r,Petro})}$}. We
estimate that this estimate is in fact a lower limit to the fraction
of UV and line emission included within the SDSS fiber aperture, based
on the compactness of the star-forming regions in the HST images and
follow-up spectra.

\begin{figure*}[t]
\begin{center}\includegraphics[width=\textwidth]{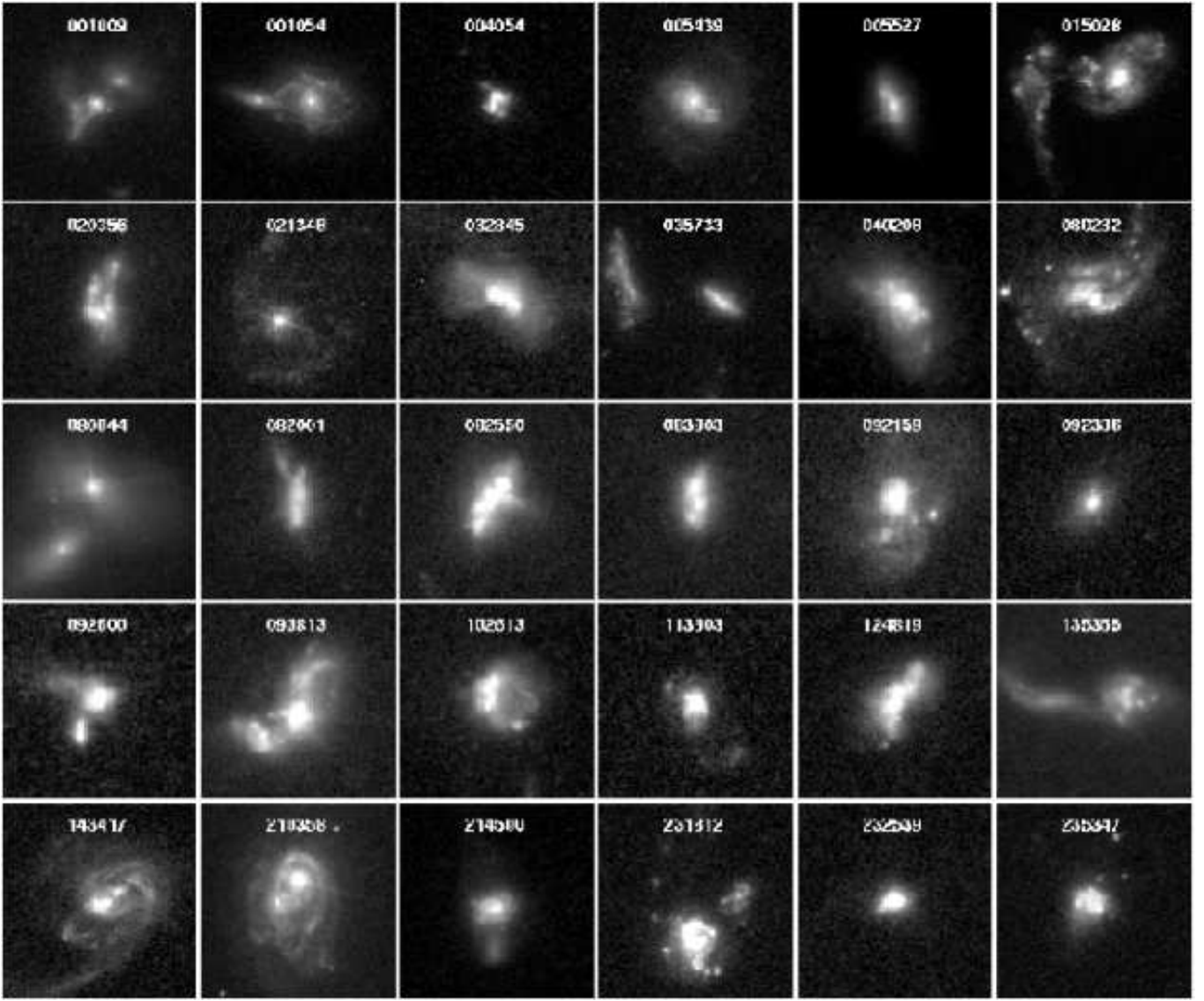}
\end{center}
\caption{\label{fig:stamps}False-color HST images of the LBA sample
  showing the (rest-frame) UV in blue/purple and optical in
  yellow/red. The images measure 6\arcsec$\times$6\arcsec. The UV
  images were rebinned (2$\times$2) to match the pixel scale of the
  optical images, and convolved with a Gaussian kernal with 0\farcs1
  (FWHM). Although most objects are highly compact in both the UV and
  optical, a small subset consists of a very bright unresolved
  component in the middle of an extended, low surface brightness
  disk. The images demonstrate a wide range of complex morphologies
  often suggestive of interactions and (post-)merging. See text for
  details.}
\end{figure*}

In Sect. \ref{sec:agn} we present a preview of data\footnote{ESO Program
082.B-0512(A)} taken
with the Fibre Large Array Multi Element Spectrograph (FLAMES) on the
European Southern Observatory (ESO) Very Large Telescope (VLT) during the night of Oct 30 2008. We used the GIRAFFE/ARGUS integral field
spectrograph (IFS) to obtain seeing-limited spectra resolving the
\ha\ and \nii$\lambda\lambda$6548\AA,6584\AA\ line complex at a
spectral resolution of $R\simeq9000$ and a spatial plate scale of
$0\farcs3$ or $0\farcs52$ pixel$^{-1}$ (depending on the seeing). Each
source was observed in 4 exposures of 400 s each. The data was
pipeline reduced and further cleaned of cosmic rays and the background
using our own routines.

More details on the LBA sample selection, observations and all
measurements used in this paper are given elsewhere \citep[Paper
  I;][]{hoopes07,basu-zych07,overzier09}. The object coordinates,
redshifts and basic quantities used in this paper are listed in Table
\ref{tab:galprops}. Emission line measurements are listed in Table
\ref{tab:lineprops}.

\subsection{Measuring Stellar Masses}

We will use stellar mass estimates that were obtained from the
SDSS/DR7 stellar mass catalogue\footnote{Available at {\tt
    http://www.mpa-garching.mpg.de/SDSS/DR7/}}. These masses are
  derived from fits to the SDSS photometry using a large grid of
  models constructed using the \citet[][BC03]{bruzual03} stellar synthesis
  library. Although the BC03 models may not include the most
  up-to-date treatment of the thermally pulsating (TP)-AGB phase, it
  has been shown that this phase mostly affects the stellar
  mass-to-light ratio in the rest-frame near-infared $H$ and $K$ bands
  \citep[e.g., see][]{eminian08}. Because near-infared data is not
  being used in the fits, we believe that our stellar mass estimates
  are relatively accurate. However, in order to further reduce the
uncertainty in these galaxy stellar mass estimates, we will derive a
``best'' mass by taking the average of the SDSS photometric mass and a
dynamical mass that we calculate from the H$\alpha$ emission line
velocity dispersion and the optical half-light radius. The two mass
estimates typically agree to within $\sim$0.3 dex with no systematic
differences (see Table \ref{tab:galprops}).

\subsection{Measuring Star Formation Rates} 
\label{sec:sfrs}

We will make use of three different estimates of the integrated star
formation rate. We use the most recent calibrations from literature as
tabulated in \citet{calzetti09}, and valid for a \citet{kroupa08}
initial mass function (IMF). The resulting SFRs are lower by a factor of
$\sim$1.5 compared to a Salpeter (1955) IMF.

\subsubsection{\ha}

SFR$_{H\alpha,0}$ is calculated from the \ha\ line luminosity and
applying a correction for dust based on the Balmer decrement
($f_{H\alpha}/f_{H\beta}$) following the recipe in
\citep{calzetti01}:
\begin{equation}
\mathrm{SFR}_{H\alpha,0}~\mathrm{[M_\odot~yr^{-1}]}=5.3\times10^{-42}L_{H\alpha,0}~\mathrm{[erg~s^{-1}]},
\end{equation}
\noindent
where $L_{H\alpha,0}=L_{H\alpha,obs}10^{0.4A_{H\alpha}}$, and
\begin{eqnarray}
A_{H\alpha}&=&k(H\alpha)E(B-V)_{gas},\\
E(B-V)_{gas}&=&\frac{\mathrm{log_{10}}[(f_{H\alpha}/f_{H\beta})/2.87]}{0.4[k(H\alpha)-k(H\beta)]},\nonumber
\end{eqnarray}
\noindent
with $k(H\alpha)=2.468$ and $k(H\alpha)-k(H\beta)=1.17$. We apply a
small correction factor of typically $\sim1.7$ due to the flux
expected outside the SDSS fiber. 

\subsubsection{\ha$+$24 $\mu$m}

An improvement to the above method is given by SFR$_{H\alpha+24}$,
which is based on a combination of the uncorrected \ha\ luminosity
($L_{H\alpha,obs}$), and the 24$\mu$m luminosity
($L_{24}\equiv\nu_{24} l_{24}$) related to the emission of dust
heated by young stars.  It has been shown that SFR$_{H\alpha+24}$ is a
robust indicator for the total SFR calibrated against Pa$\alpha$
measurements in a large sample of nearby galaxies \citep{calzetti07}:
\begin{eqnarray}
\mathrm{SFR}_{H\alpha+24}~\mathrm{[M_\odot~yr^{-1}]}&=&\\
5.3\times10^{-42}&\times&(L_{H\alpha,obs}+0.031L_{24})~\mathrm{[erg~s^{-1}]}.\nonumber
\end{eqnarray}
\noindent
As the exact, individual infrared spectral energy distributions are
currently uncertain, we choose not to apply any $K$-corrections to the
observed 24$\mu$m fluxes. Based on the average corrections found for a
collection of spectral templates presented by \citet[][see their
  Figure 9]{rieke08} we estimate that the actual SFRs could be between 0.0
and 0.3 dex higher depending on the redshift, spectral shape, and total IR luminosity.

\subsubsection{FUV}

We will make use of SFR$_{FUV,0}$, which is based on the
dust-corrected far-UV luminosity using the empirical correlation between
the attenuation of the UV and the IR emission from dust heated by
obscured stars \citep[the ``$\beta$--$IRX$
  relation'',][]{meurer95}. The correlation has been shown to work well for relatively unobscured
star-forming and starburst galaxies at both low and high redshift
\citep[e.g.][]{meurer95,meurer99,seibert05,reddy06,salim07},
but it must be noted that the situation is more complex for galaxies
that have more complicated star formation histories
\citep[e.g.][]{kong04,johnson07}, very young ages
\citep[e.g.][]{reddy06,siana08} or that are heavily obscured
\citep[e.g.][]{meurer99,goldader02,reddy06}.
We use the calibration between UV colour and
$A_{FUV}$ from \citet{treyer07}, $A_{FUV}=4.05(m_{FUV}-m_{NUV})-0.18$
mag, finding:
\begin{eqnarray}
\mathrm{SFR}_{FUV}~\mathrm{[M_\odot~yr^{-1}]}&=&\\
8.1\times10^{-29}&\times& l_{1500\AA}10^{0.4A_{FUV}}~\mathrm{[erg~s^{-1}~Hz^{-1}]}.\nonumber\\
\end{eqnarray}
\noindent
Using the spectral slope $\beta=2.32\mathrm{log}_{10}(m_{FUV}-m_{NUV})-2.0$ with
$f_\lambda\propto\lambda^\beta$ we further applied a small $K$-correction to obtain the monochromatic luminosity at
rest-frame 1500\AA, $l_{1500\AA}$.

\begin{figure}[t]
\begin{center}
\includegraphics[width=\columnwidth]{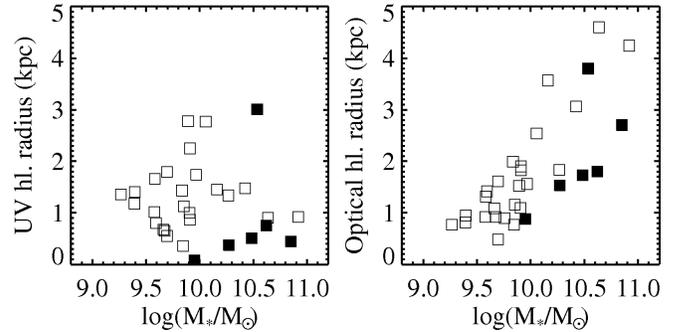}
\end{center}
\caption{\label{fig:rvsm}Stellar mass versus the half-light radius
  measured in the UV ({\it left panel}) and optical ({\it right
    panel}) images.  Solid squares mark the subset of DCOs, in
  anticipation of results obtained in Section \ref{sec:cusps}.}
\end{figure}

\section{Results}

\subsection{UV--optical morphologies}

In Fig. \ref{fig:stamps} we show 6\arcsec$\times$6\arcsec\ UV-optical
false-color postage stamps of 30 LBAs, including the 8 objects studied
in Paper I. In all cases, both the UV and the optical light is
dominated by the intense light from young massive stars in very
compact starburst regions.  The optical images furthermore allow us to
see the faint emission from older stars displaying a wide range of
(often complex) morphologies, which include clumpy or chain-like
galaxies, interacting pairs, a ring galaxy, and perturbed (spiral)
disks.  In Fig. \ref{fig:rvsm} we plot the half-light radii versus the
stellar masses. The global median half-light radii of the galaxies are
$\approx$1 kpc and $\approx$1.5 kpc in the UV and optical
respectively. The galaxy mass estimates range from a few times $10^9$
to $\lesssim10^{11}$ $M_\odot$ (Table \ref{tab:galprops}). The sizes
and the stellar masses of these galaxies correspond well to those of
typical Lyman Break Galaxies \citep[e.g.][]{shapley01,papovich01,giavalisco02,ferguson04,stark09}.

\begin{figure}[t]
\begin{center}
\includegraphics[width=\columnwidth]{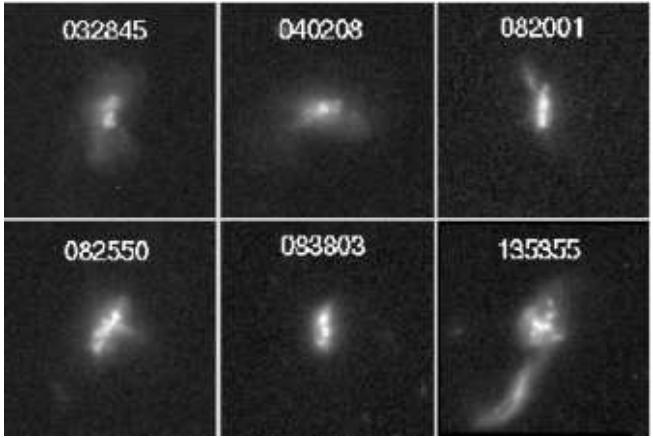}
\end{center}
\caption{\label{fig:clumps}Illustration of HST optical morphologies of 
typical LBAs having several star-forming clumps of comparable brightness. These are
to be contrasted with the six LBAs having a dominant compact central
object shown in Fig. \ref{fig:points}. The images measure 10\arcsec$\times$10\arcsec.}
\end{figure}

\begin{figure}[t]
\begin{center}
\includegraphics[width=\columnwidth]{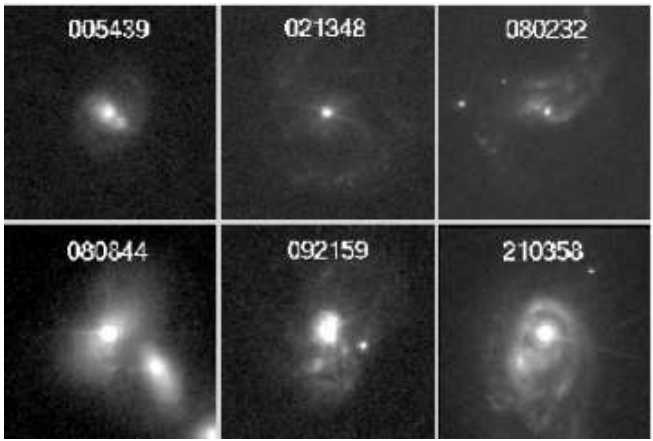}
\end{center}
\caption{\label{fig:points}HST optical morphologies of the six LBAs
  having a dominant compact central object. The images measure 10\arcsec$\times$10\arcsec.}
\end{figure}

\begin{figure*}[t] 
\begin{center} \includegraphics[width=0.8\textwidth]{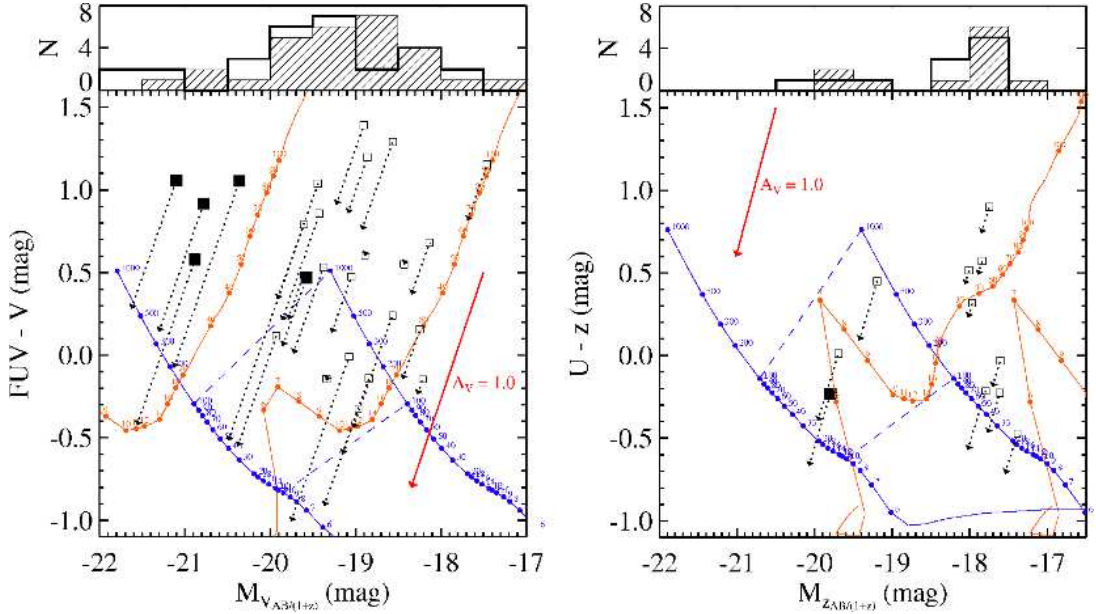} 
\end{center} 
\caption{\label{fig:cc}{\it Bottom panel:} Color-magnitude diagram of
  star-forming clumps identified in the HST WFPC2 images.  Solid
  points show measured values, while dotted lines connecting to open
  points indicate dust-corrected values based on the Balmer decrement
  and assuming a constant relation between reddening of the gas and
  the stellar continuum following \citet{calzetti01}. Coloured lines
  indicate population synthesis tracks modeled using Starburst99 for
  either constant bursts with a SFR of 1 and 10 $M_\odot$ yr$^{-1}$
  (right and left blue tracks) or an instantaneous burst normalized to
  a total stellar mass of $10^8$ $M_\odot$ and $10^9$ $M_\odot$ (right
  and left red tracks). Ages in Myr have been indicated along the
  tracks. Blue dashed lines indicate the ages at which the constant
  burst models reach a total stellar mass of $10^8$ $M_\odot$ and
  $10^9$ $M_\odot$.  The models used have a LMC metallicity, include
  the nebular continuum, and were redshifted to $z=0.15$. The red
  arrow indicates the reddening vector corresponding to an extinction
  of $A_V=1.0$ mag. The six DCOs are indicated by the filled squares.
  {\it Top panel:} Histogram of the absolute magnitude distribution
  corresponding to the clumps shown in the bottom panel. The measured
  (dust-corrected) distribution is indicated by the shaded (open)
  histogram. Typical massive star clusters in nearby galaxies have
  $M_V\simeq-(16-14)$ that would be as bright as $M_V\simeq-(19-17)$ when
  observed at the young age of $\sim$10 Myr. The LBA clumps, and the
  DCOs in particular, are generally much brighter than typical massive
  star clusters. See Sect. \ref{sec:clumps} for discussion.}
\end{figure*}

As noted already in Paper I, a significant fraction of both the UV and
optical emission is seen to come from luminous, very compact
clumps. The much larger size of the sample presented here allows us to
investigate this in more detail. In Figs. \ref{fig:clumps} and
\ref{fig:points} we show some close-ups of the rest-frame optical HST
images at high contrast.  LBAs typically consist of a collection of
centrally-located compact clumps having a range in magnitudes (e.g.,
Fig. \ref{fig:clumps}).  However, in a minority of cases the LBA
contains a single, dominant luminous point-like source at or near the
center of the galaxy (i.e., 005439, 021348, 080232, 080844, 092159,
and 210358). In these sources, the central source typically
  contains $\sim$50\% (20 and 77\% being the lowest and the highest,
  respectively) of all the UV light. In the optical, the contribution
  from the host galaxies is larger, somewhat reducing the light
  fraction of the central source to $\sim20-30$\%. Here we have used
  an aperture of 0\farcs2 radius for the central point-like source and
  an aperture of 10\arcsec\ to estimate the total light of the
  galaxy. Because of their visual prominence and in anticipation of
some of their remarkable properties that will be investigated in the
following sections, we will
henceforth refer to this subset of the clumps as ``dominant central
objects'' (DCOs).  A collage of images of the six objects hosting DCOs
is shown in Fig. \ref{fig:points}.

\subsection{Color-magnitude diagrams: estimated clump masses}

We have measured the magnitudes and colors of each of the clumps and
DCOs identified by eye inside circular $0\farcs4$ diameter
apertures. The backgrounds were determined inside a local annulus and
subtracted. We did not make a correction for the (small) amount of
flux expected to fall outside the aperture due to the HST PSF.  
In Fig. \ref{fig:cc} we show the color-magnitude diagrams
for all objects for which two filters were available, either $FUV$ and
$V$ (shown on the left) or $U$ and $z$ (shown on the right). Open and
filled squares indicate the measured values, while the dotted lines indicate the
dust-corrected values based on the Balmer decrement
($f_{H\alpha}$/$f_{H\beta}$) using a constant relation between
reddening of the gas and the stellar
continuum\footnote{$E(B-V)_*=0.44E(B-V)_{gas}$}
\citep{calzetti01}. Unfortunately we need to assume that the Balmer decrement as measured
from the integrated SDSS spectrum applies to the much smaller clumps
as well, although this is not necessarily correct.    
The results are tabulated in Table
\ref{tab:clumpprops}. The clumps are very luminous with a median
absolute magnitude in $V$ (approximately rest-frame $B$) of
$\simeq-20$ mag, and very blue ($FUV-V\simeq0.0$). The DCOs are the
most luminous having $M_B \sim$ --20 to --22 mag. The small gap in the
magnitude histograms in the top panels of Fig. \ref{fig:cc} may suggest
that the DCOs are not just a bright tail of the general clumps
distribution, but perhaps a distinct class of object.

\begin{figure}[t] 
\begin{center}
\includegraphics[width=\columnwidth]{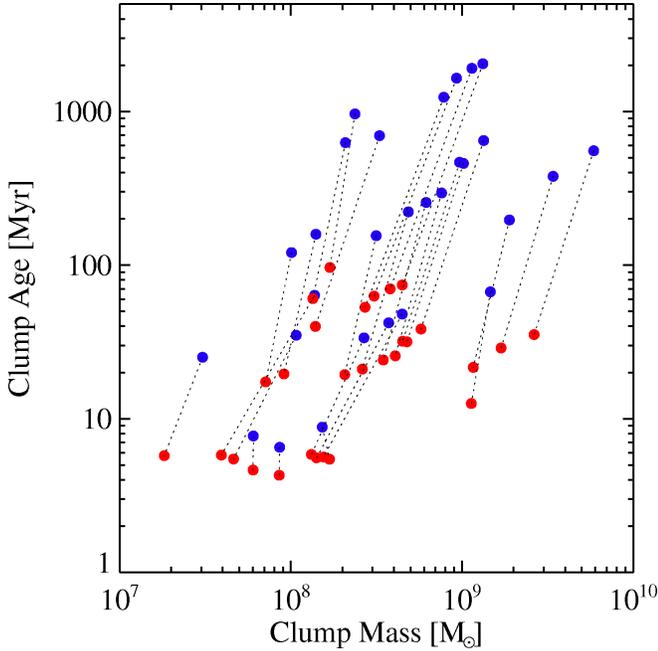} 
\end{center} 
\caption{\label{fig:sfh}Clump masses and ages estimated from
  the color-magnitude diagrams presented in Fig. \ref{fig:cc}, assuming  
 either an instantaneous star formation history (red points) or a
 continuous star formation history (blue points). The masses estimated
 using the continuous model are only $\sim2\times$ higher on
average compared to those derived for instantaneous bursts, while the
ages estimated using the continuous model are significantly higher
compared to the instantaneous model.}
\end{figure}

A comparison with Starburst99 5.1 \citep{vazquez05} stellar synthesis
models indicates that the stellar masses of the clumps shown in
Fig. \ref{fig:cc} are typically a few times $10^7$ to a few times
$10^{8}$ $M_\odot$, while the DCOs have typical masses near or in
excess of $10^{9}$ $M_\odot$ (see Table \ref{tab:clumpprops}).  It is
important to note that these mass estimates are relatively insensitive
to dust extinction, since the dust extinction vector runs nearly
parallel to the age vector in these plots. As shown in
Fig. \ref{fig:sfh}, the masses are 
relatively insensitive to the star formation history (SFH) that we
adopt. The figure also illustrates that it is much
less straightforward to determine the typical stellar age of the
clumps without accurate information about the extinction and SFH. For
our adopted extinction corrections, the instantaneous burst models
imply ages in the range from $\sim$5 to 100 Myr (red points in Fig. \ref{fig:sfh}). Alternatively, if we
adopt a continuous SFH the ages would be substantially higher, in the
range $\sim$100--1000 Myr (blue points), while the SFRs would have to be on the
order of $\sim$10 $M_\odot$ yr$^{-1}$ in order to explain the high
luminosity of the brightest clumps (left blue track in
Fig. \ref{fig:cc}). The masses would be $\sim2\times$ higher on
average compared to those derived for instantaneous bursts (Table
\ref{tab:clumpprops}). However, there are several arguments to make in
favor of an instantaneous SFH (or a SFH having a very short
$e$-folding time). We note that at least three of the sources show
prominent Wolf-Rayet features indicating the presence of very massive
stars at a special phase only $\sim$3--4 Myr after the onset of a
massive burst \citep[e.g.][]{brinchmann08b}. This would favor the
instantaneous burst model. Also, we note that the very compact sizes of
these objects (and thus the short dynamical times) are consistent with
a short formation time ($t_{form} \gtrsim t_{dyn}$).

\subsection{Clump sizes}
\label{sec:sizes}

It is clear from simple visual inspection of the HST images that the
DCOs are marginally-resolved at best (note the strong diffraction
spikes associated with the DCOs in Fig. 3). Many of the other clumps
appear either as faint point-like sources or larger spatially resolved
regions (see examples in Fig. \ref{fig:clumps}).

\begin{figure*}[t]
\begin{center}
\includegraphics[width=0.292\textwidth]{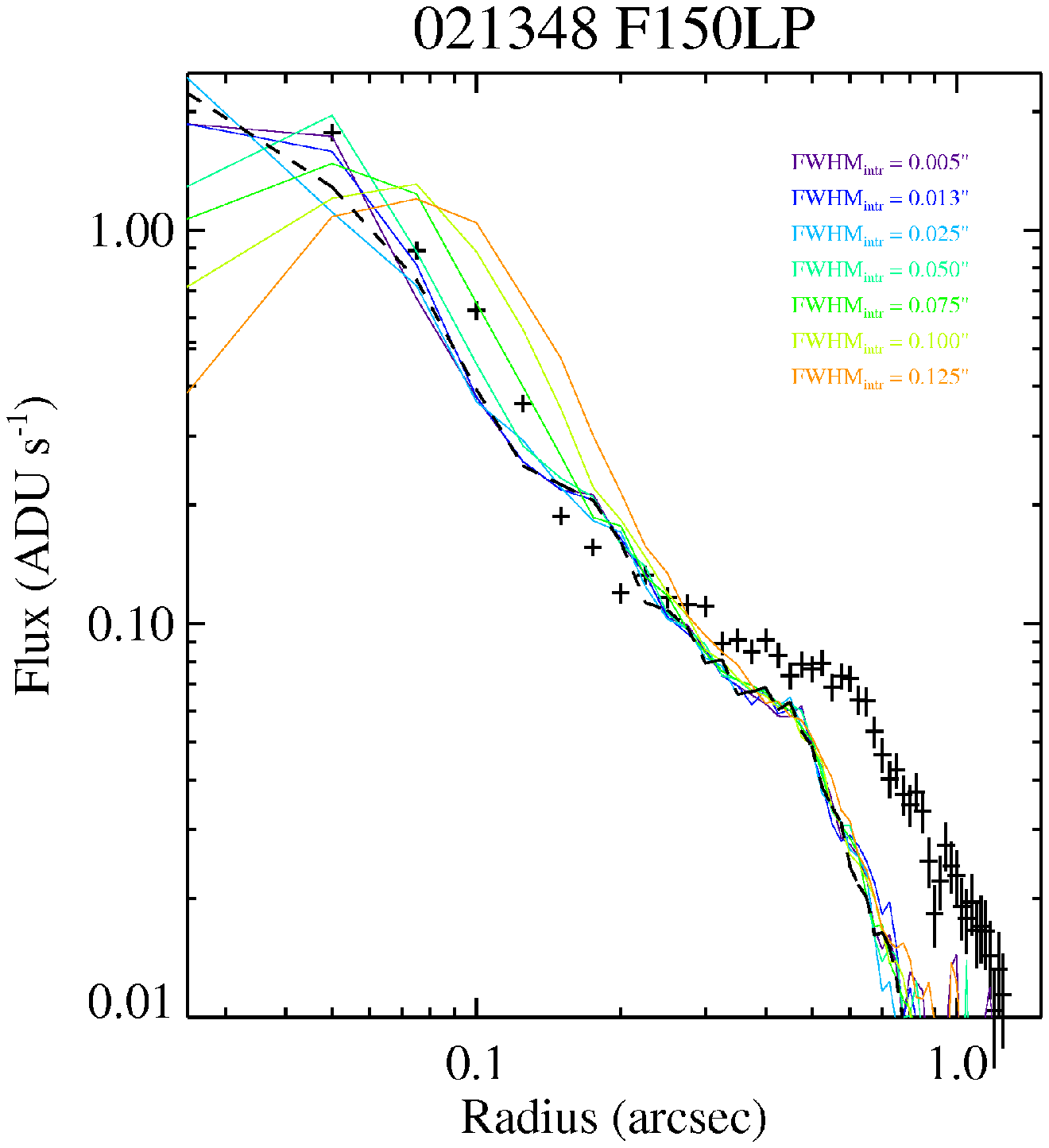}
\includegraphics[width=0.3\textwidth]{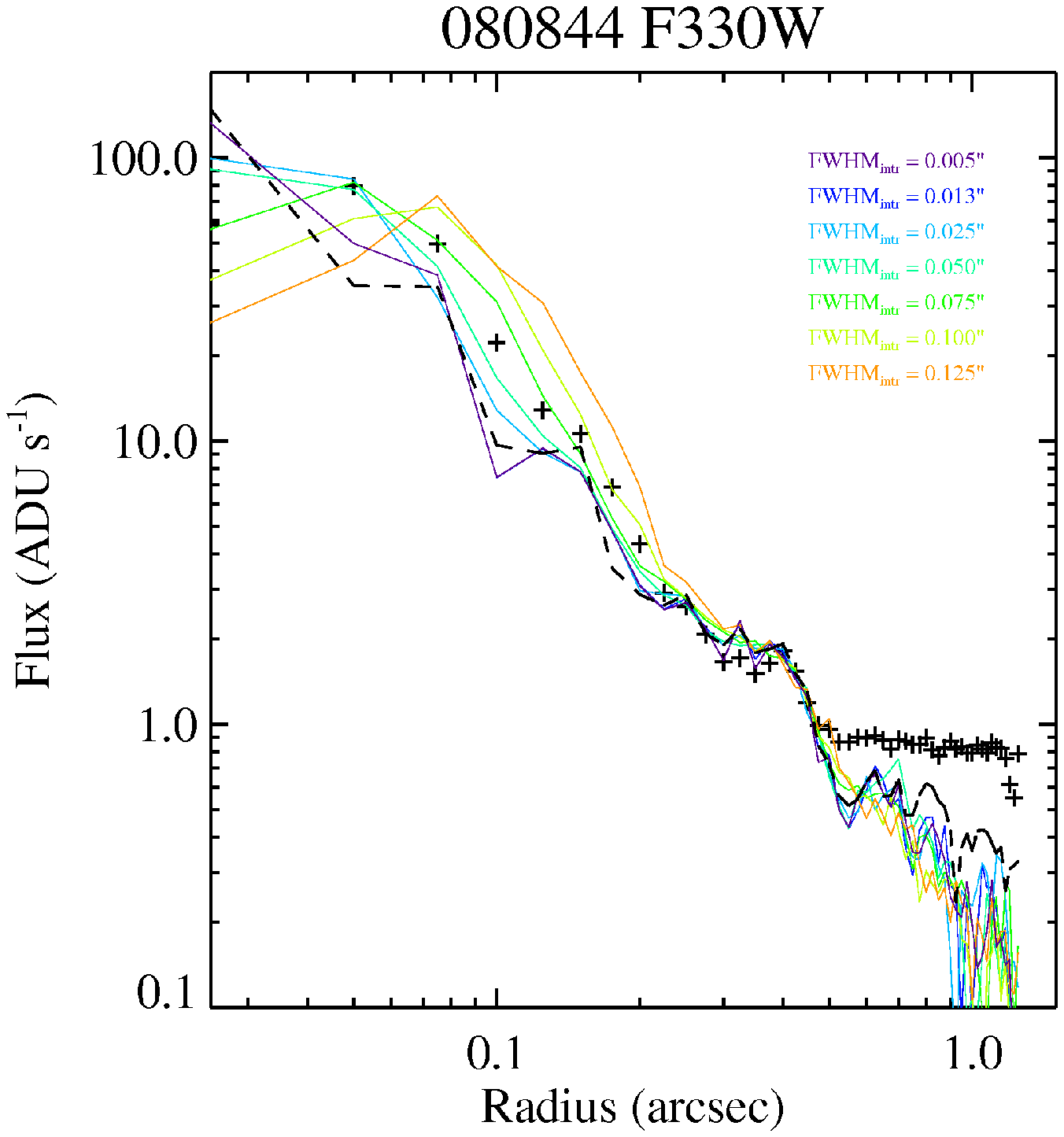}
\includegraphics[width=0.3\textwidth]{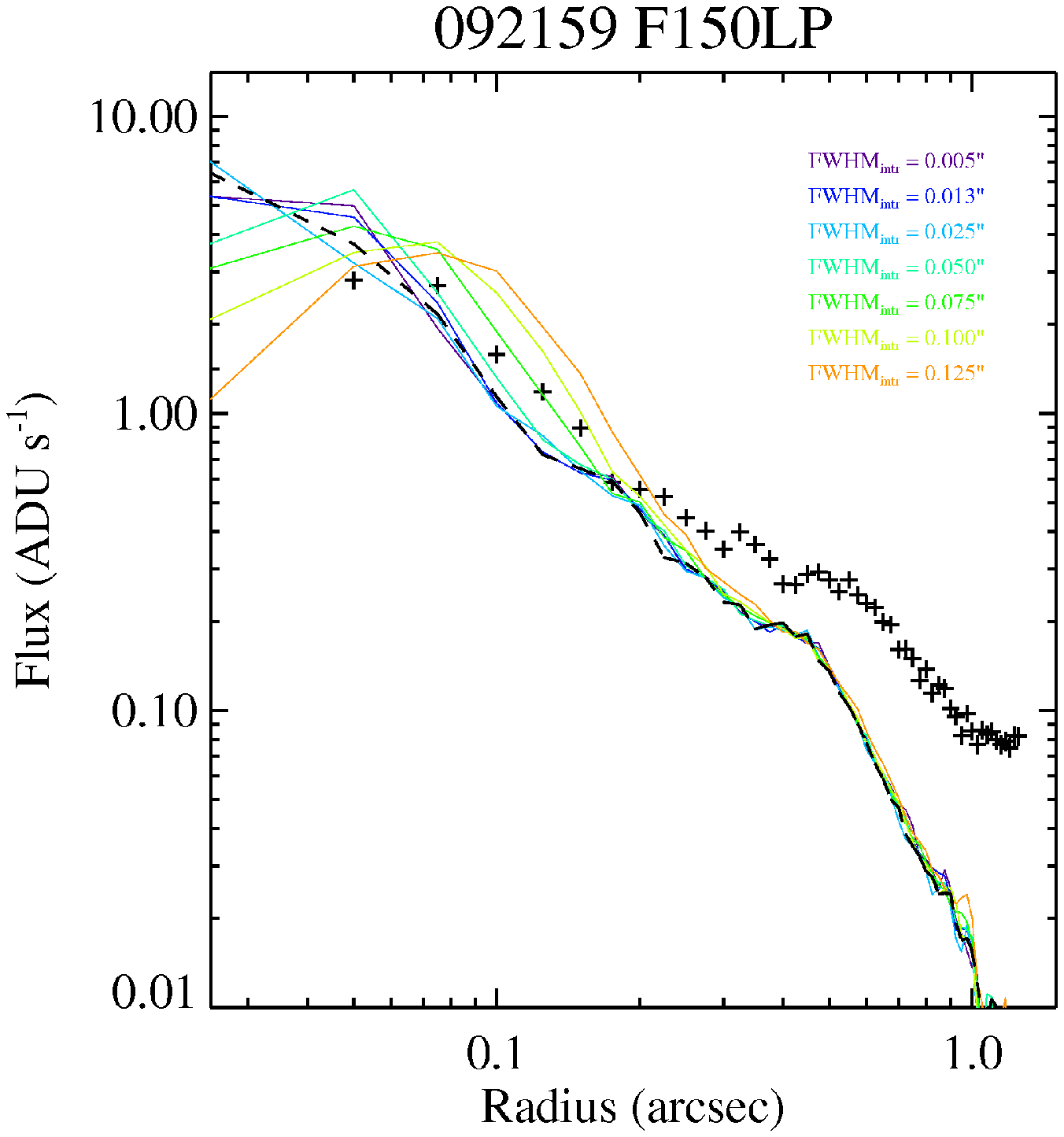}
\end{center}
\caption{\label{fig:psf}Radial flux profiles of the three strongest,
  point-like sources identified in the HST images. Panels show the
  measured count rates ({\it plusses}) in either the ACS/SBC or HRC
  images compared to the measured profile of a model PSF ({\it dashed
    line}) modeled using the TinyTim 3.0 software \citep{krist95}. To
  investigate the size limits we can place on the nuclear sources, we
  simulated a range of 2D Gaussian models having a FWHM of
  $0\farcs005$--$0\farcs125$ and convolved them with the model
  PSF. The results are shown using coloured lines. In all three cases,
  the data is most consistent with an object no larger than
  $\approx0\farcs075$ (FWHM; corresponding to 135--320 pc at
  $z=0.1-0.3$).}
\end{figure*}

To obtain the best constraints on the sizes of the DCOs, we have first
examined the radial flux profile of the three brightest examples,
using the ACS UV images as they have smaller pixels and a sharper PSF
compared to the optical images. The result is shown in
Fig. \ref{fig:psf}. Each panel shows the measured count rate ({\it
  plusses}) compared to the measured profile of a model PSF ({\it
  dashed line}) modeled using the TinyTim 3.0 software
\citep{krist95}. We also simulated a range of 2D Gaussian models
having a FWHM of $0\farcs005$--$0\farcs125$ and convolved them with
the model PSF. The results are shown using coloured lines. In all
three cases, the data is most consistent with an object no larger than
$\approx0\farcs075$ (FWHM), corresponding to effective radii no larger than
$\simeq$70--160 pc at $z=0.1-0.3$).

\begin{figure}[t]
\begin{center}
\includegraphics[width=\columnwidth]{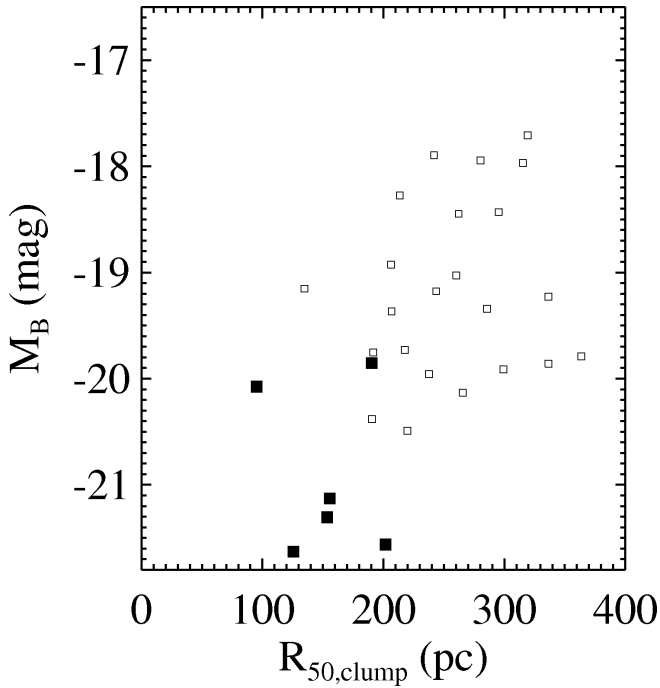}
\end{center}
\caption{\label{fig:reff}Absolute magnitude versus the optical half-light
  radius of the brightest clump in each galaxy estimated in
  Sect. \ref{sec:sizes}. The
  six dominant central objects (DCOs) are indicated by filled
  squares. }
\end{figure}

In order to be consistent across our entire clump sample and with our
earlier definition of the clumps, we have measured the physical sizes
of the brightest clumps in each LBA by measuring the optical
half-light radii of each clump within the same fixed angular apertures
used in Fig. \ref{fig:cc}. The radii were deconvolved assuming a PSF
of $0\farcs1$ (FWHM).  In Fig. \ref{fig:reff} we plot the clump radii
versus the extinction-corrected clump absolute magnitudes.  The
typical clumps have radii of $\sim$200 to 400 pc, while the DCOs have
radii of $\sim$100-200 pc. Comparing the sizes obtained here with
those obtained for the three DCOs using our more careful model
simulations shown in Fig. \ref{fig:psf} suggests that the more crude
analysis may somewhat overestimate the true sizes of at least the DCOs.    
In any case, Fig. \ref{fig:reff} shows an inverse correlation
between clump luminosity and size, largely defined by the extreme
properties of the DCOs. Their high luminosities (corresponding to very
high stellar masses) and compact sizes imply very high densities. We
will discuss the implications of this remarkable result later in the
paper.

\subsection{The relation between clumps and their host galaxies}
\label{sec:panels}

\begin{figure*}[t]
\begin{center}
\includegraphics[width=0.7\textwidth]{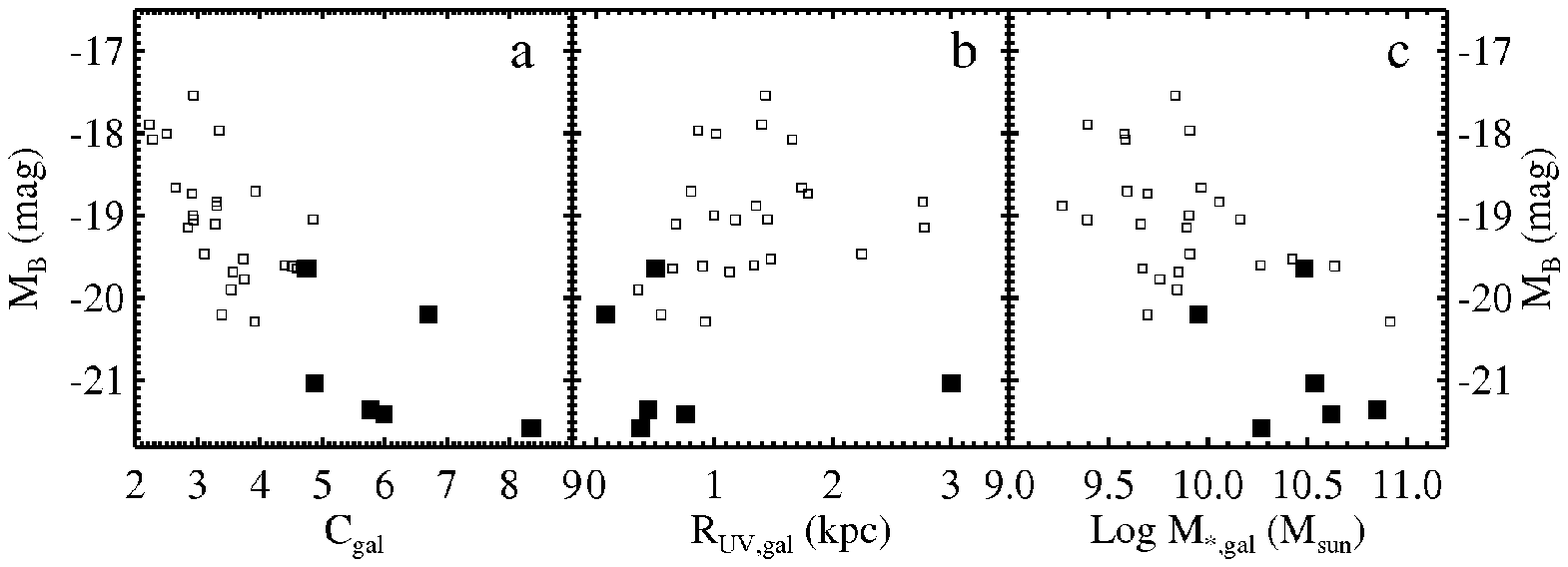}
\includegraphics[width=0.7\textwidth]{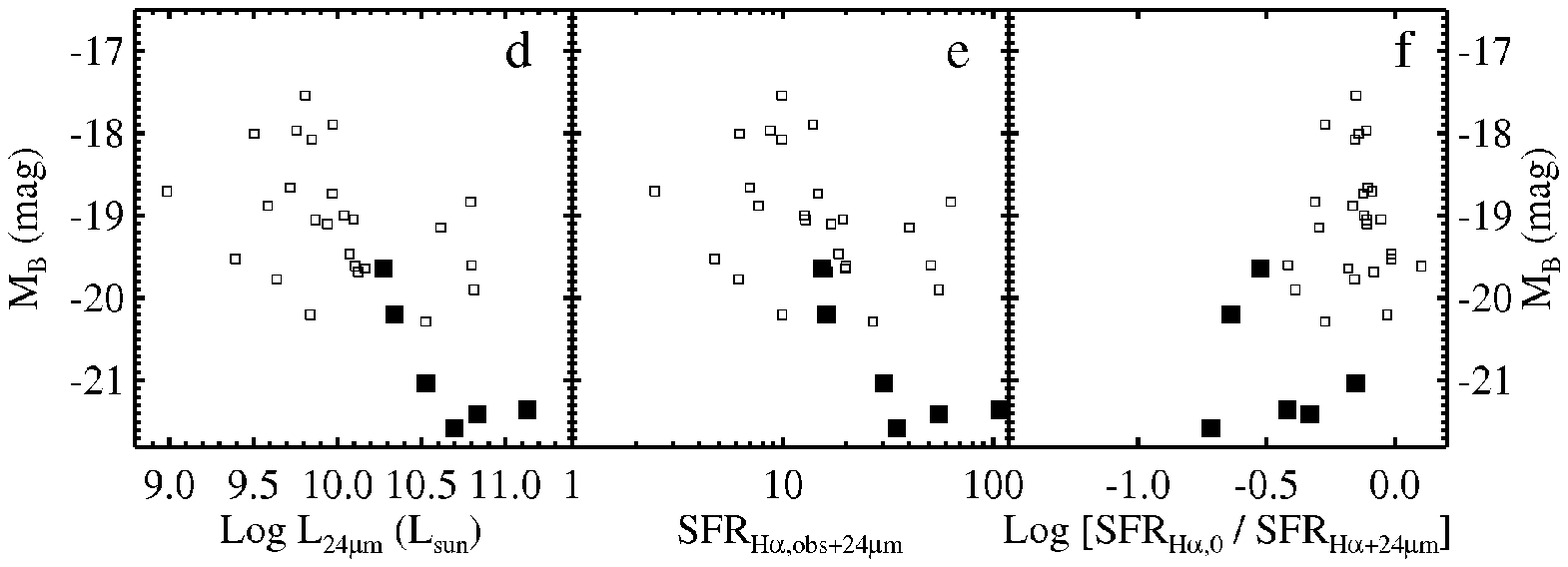}
\end{center}
\caption{\label{fig:panels1} Correlations between the absolute
  $B$-band magnitudes ($M_B$) of the star-forming clumps and various
  properties of the host galaxies measured from the HST optical image,
  the SDSS spectrum and the Spitzer IR photometry. The six dominant
  central objects (DCOs) are indicated by filled squares. Panels show
  {\it (a)} $M_B$ versus optical concentration index, {\it (b)} galaxy
  UV half-light radius, {\it (c)} stellar mass, {\it (d)} 24$\mu$m
  luminosity, {\it (e)} total star formation rate, and {\it (f)} the
  ratio of the \ha\ SFR, SFR$_{H\alpha,0}$, corrected for dust using
  the Balmer lines to SFR$_{H\alpha,obs+24{\mu}m}$, the SFR based on
  the attenuated \ha\ corrected for dust using the 24$\mu$m luminosity
  and \citet{calzetti07}. We observe the following trends. The
  brightest clumps tend to have {\it (a)} larger host concentrations
  $C$, {\it (b)} smaller effective radii, {\it (c)} larger host
  stellar masses $M_*$, {\it (d)} larger host mid-IR luminosities,
  {\it (e)} larger host star formation rates, and {(\it f )}
  relatively low total SFRs based on \ha\ corrected for dust using the
  Balmer decrement compared to SFRs based on \ha\ corrected for dust
  based on the 24$\mu$m emission from dust heated by young stars. The
  six dominant central objects (DCOs) are indicated by filled
  squares. The DCO having a relatively large UV half-light radius of
  $\sim$3 kpc seen in panel {\it b} is object 080232. See \S\ref{sec:panels} for details.}
\end{figure*} 

We have demonstrated above that massive star-forming clumps are an
essential ingredient in our sample of local Lyman Break Galaxy
analogs, and have found that a subset contain a single extremely
massive (up to several $\times 10^9 M_{\odot}$) compact (radius $\sim
10^2$ pc) dominant central object (DCO). How do the properties of
these compact structures relate to the overall properties of the host
galaxies?

The results are shown in the six panels of Fig. \ref{fig:panels1}. In
all panels we plot the extinction-corrected absolute $B$-band
magnitudes of the clumps measured within a fixed radius of 500 pc, in
order to account for the effect that the angular sizes change by a
factor of $\sim$2.5 over the redshift range $z=0.1-0.3$. The results
are as follows.

\smallskip
\noindent {\it Galaxy Concentration.--} In panel {\it (a)} we plot
clump $M_B$ versus the optical concentration parameter\footnote{To
  calculate the concentration we follow the
procedures described in full detail in \citet{lotz06} and Paper I. First,
we use SExtractor to make an object segmentation map and mask out
neighboring objects. The image is background subtracted, and we
calculate an initial Petrosian radius ($r_P$ with $\eta\equiv0.2$)
using the object center and (elliptical) shape information from
SExtractor. We then smooth the image by $\sigma=r_P/5$ and create a
new segmentation map by selecting those pixels that have a surface
brightness higher than the mean surface brightness at the Petrosian
radius. We recalculate the object center by minimizing the second
order moment of the flux, and then recalculate the Petrosian radius in
the original image using this center. The total flux is defined as the
flux within a radius of $1.5\times r_P$ and $C$ is then calculated in
circular apertures containing 20 and 80\% of the light.} of the host
galaxy, $C\equiv5$log$\large(\frac{R_{80}}{R_{20}}\large)$. In general the more luminous clumps occur in the LBAs
with larger concentrations. The six dominant central objects (filled
squares) are the most extreme. Inspection of Fig. \ref{fig:stamps}
shows that these objects are indeed systematically found inside
larger, optically faint disks compared to the sample on average.
\smallskip

\noindent{\it Galaxy Half-light radius.--} In panel {\it (b)} we plot
$M_B$ versus the half-light radii for each galaxy measured in the UV
images. This plot shows that the total UV light in galaxies hosting a
DCO is dominated by the DCO itself, albeit with one notable
  exception (DCO object 080232 having $R_{UV}\sim3$ kpc).

\smallskip
\noindent {\it Stellar mass.--} In panel {\it (c)} we plot $M_B$
versus the stellar mass of the hosts. The brighter clumps are
predominantly found in the more massive objects in our sample. The
DCOs are found in galaxies with masses $M_*\simeq10^{10-11}$
$M_\odot$.

\begin{figure}[t]
\begin{center}
\includegraphics[width=\columnwidth]{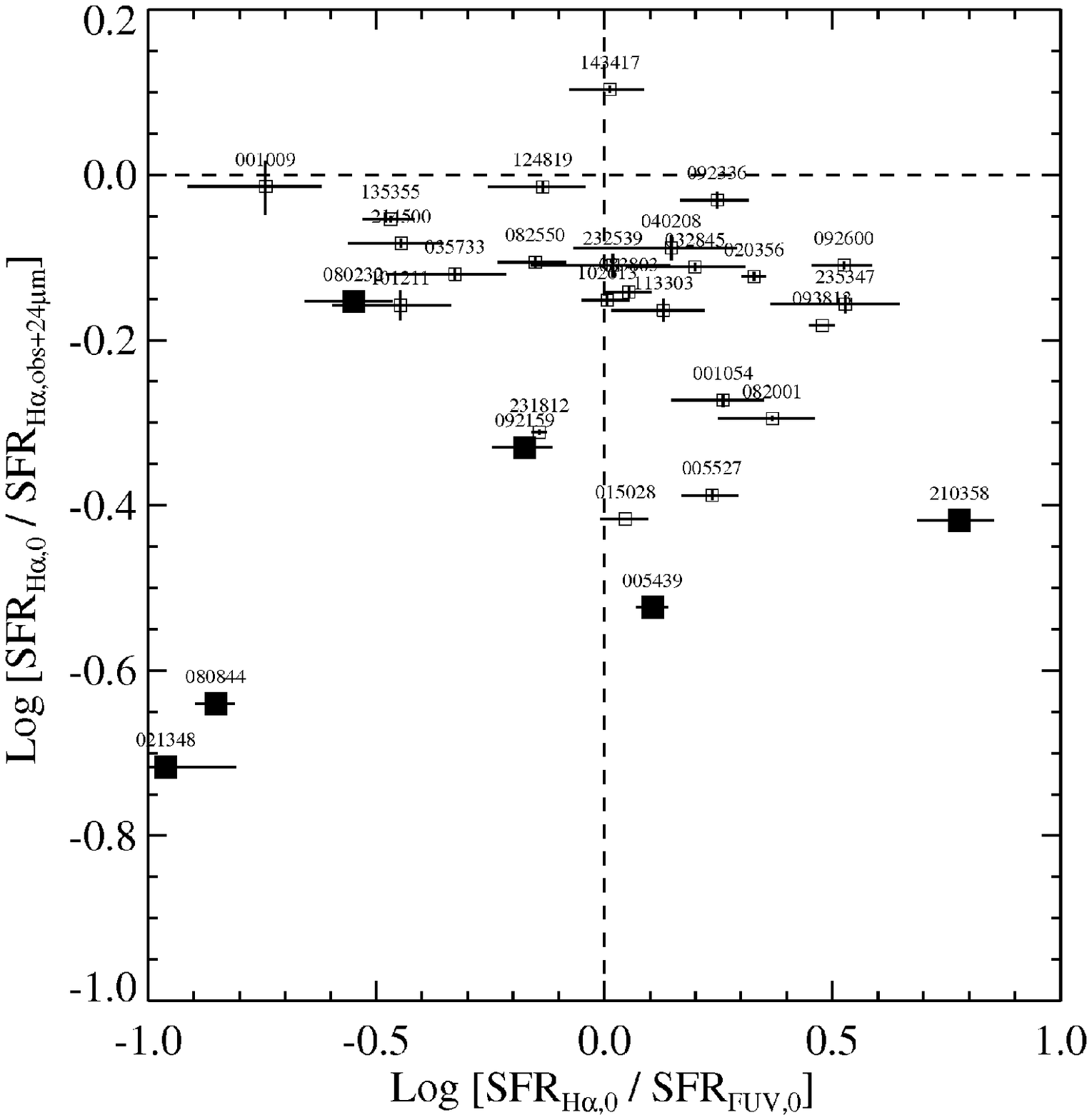}
\end{center}
\caption{\label{fig:sfrs} The ratio of SFR$_{H\alpha,0}$/SFR$_{FUV,0}$
  versus SFR$_{H\alpha,0}$/SFR$_{H\alpha,obs+24{\mu}m}$ of clump host
  galaxies. The SFRs were calculated using the recipes given in
  Sect. \ref{sec:sfrs}.  Under reasonably standard assumptions for
  young starburst galaxies, all three SFR indicators are expected to
  give more or less consistent answers, and we expect the sources to
  line up on the dashed lines. We find significant outliers and
  discuss the implications in Sections \ref{sec:agn} and
  \ref{sec:lines}. The DCOs are marked by filled squares.}
\end{figure}

\smallskip
\noindent {\it Mid-IR luminosity.--} In panel {\it (d)} we show the
relation between optical clump luminosity and the host galaxy
luminosity at 24$\mu$m. We find a correlation with $L_{24{\mu}m}$
indicating that the galaxies hosting the brighter clumps seen in the
UV--optical are also the stronger IR sources. Even though selected on
the basis of their high luminosity in the far-UV, the galaxies with
DCOs have mid-IR luminosities similar to those of luminous infrared
galaxies (LIRGs).

\smallskip
\noindent {\it SFR.--} These high mid-IR luminosities imply high
star-formation rates (SFR$_{H\alpha+24}$), as shown in panel {\it
  (e)}.  The star formation rates are systematically larger in the
LBAs with the more luminous clumps, and the six galaxies with DCOs
have star formation rates ranging from 15 to nearly 100 $M_{\odot}$
per year.

\smallskip
\noindent {\it Comparing SFR indicators.--} In panel {\it (f)} we plot
$M_B$ versus the ratios of the two commonly used SFR indicators
SFR$_{H\alpha}$ and SFR$_{H\alpha,obs+24}$.  For the LBAs with the
lower luminosity clumps, the majority show good agreement between the
two SFR indicators. The LBAs with the brighter clumps are
sytematically displaced towards low values of
SFR$_{H\alpha,0}$/SFR$_{H\alpha,obs+24}$, with the galaxies having a
DCO being among the most extreme.
To further investigate the discrepancy between SFR$_{H\alpha,0}$ and
SFR$_{H\alpha,ob+24}$ for the brightest sources, we compare
SFR$_{H\alpha,0}$/SFR$_{H\alpha,obs+24}$ with the ratio
SFR$_{H\alpha,0}$/SFR$_{FUV,0}$. The result is shown in
Fig. \ref{fig:sfrs} comparing the three SFR indicators. For the
majority of sources all three indicators agree to within a factor of
$\sim$ 2 to 3, but there are outliers for which the SFRs derived from
the extinction-corrected H$\alpha$ luminosity are significantly lower
than either both or one of the other two SFRs.  In other words, the
ratio of the dust-corrected H$\alpha$ luminosity relative to the
luminosities of the mid-IR and (intrinsic) far-UV continua is
systematically small in the galaxies with DCOs compared to normal star
forming galaxies and the less extreme LBAs. We will refer to these
offsets in Sections \ref{sec:agn} and \ref{sec:lines} below.

\begin{figure}[t]
\begin{center}
\includegraphics[width=\columnwidth]{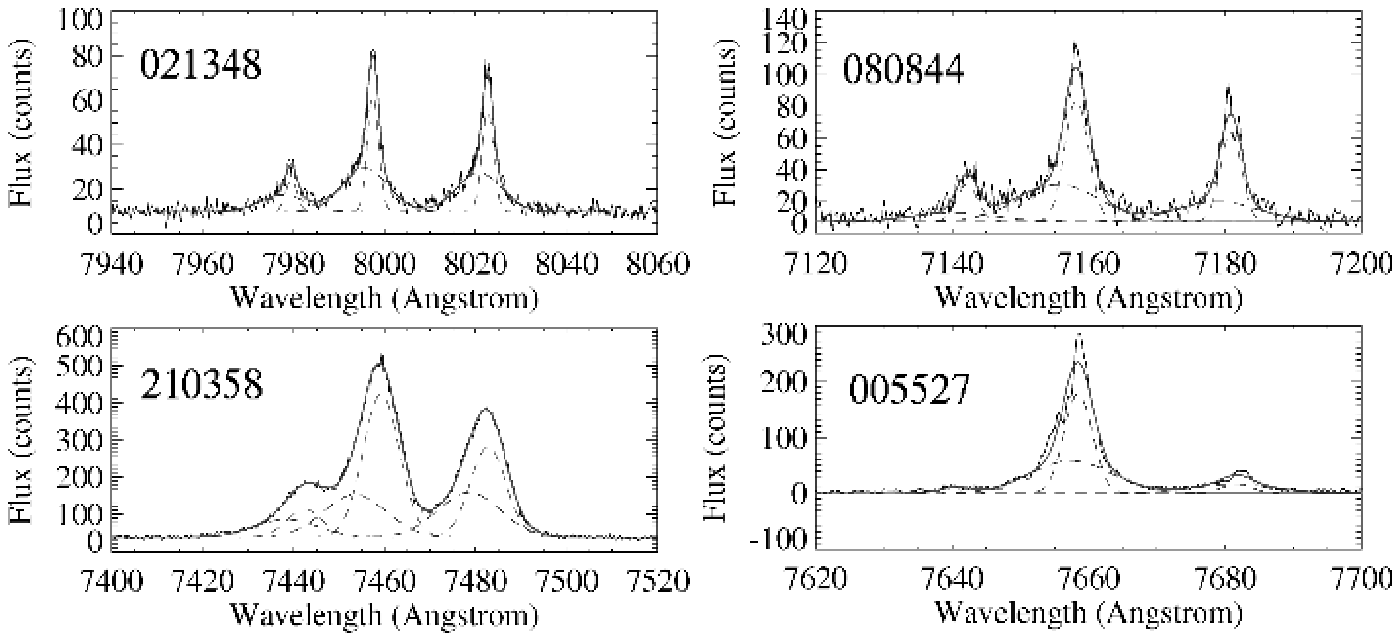}
\end{center}
\caption{\label{fig:haspec}FLAMES/GIRAFFE/ARGUS spectra showing the
  \ha\ and \nii$\lambda\lambda$6548\AA,6584\AA\ line complex within a
  small aperture placed on the compact clumps in four of the
  sources. Sources 021348, 080844 and 210358 host the most luminous
  and compact clumps seen in the LBA sample and all lie in the SF/AGN
  composite region of the BPT diagram making them candidates for
  hosting a (weak) AGN. Source 005527 (bottom right panel) lies in the
  \hii-region dominated part of the BPT diagram, eventhough it has one
  of the largest BPT offsets $d_P$ (see Fig. \ref{fig:bpt} and panel
  (a) in Fig. \ref{fig:panels1}). The line profiles were fitted with a
  series of narrow and broad Gaussian components indicated by red and
  blue dashed lines, respectively. The fits show that each source has
  significant line asymmetries on the blue side of the lines that are
  consistent with a relatively broad,
blueshifted component (Overzier et al. in preperation).  Such blue
asymmetries are typical for galactic outflows in starburst galaxies at
both low and high redshift. See Sect. \ref{sec:agn} for details.}
\end{figure}

\begin{figure*}[t]
\begin{center}
\includegraphics[width=0.7\textwidth]{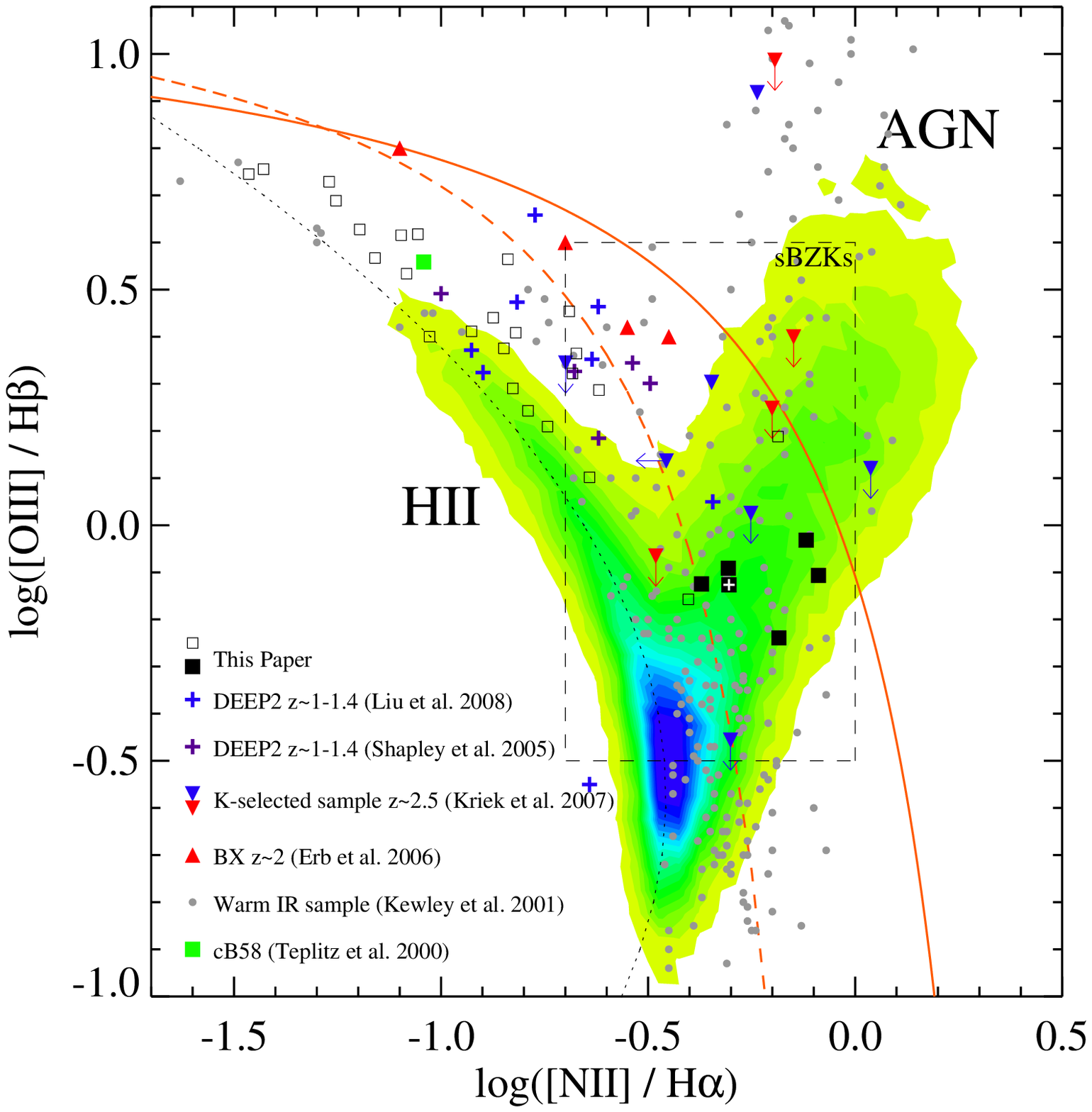}
\end{center}
\caption{\label{fig:bpt}BPT diagram of the galaxies in our sample
(open/filled squares). Filled squares correspond to the LBAs 
  hosting the DCOs. Comparison data shown are samples of star-forming
  galaxies at high redshift (green square: LBG cB58 \citep{teplitz00};
  plusses: DEEP2 galaxies at $z\sim1$ \citep{shapley05b,liu08}; red
  triangles: ``BX'' galaxies at $z\sim2$ \citep{erb06a}; blue (red) upside-down
  triangles: star-forming (quiescent) $K$-selected galaxies at
  $z\sim2.5$ \citep{kriek07}; dashed box region: star-forming ``s$BzK$'' 
galaxies at $z\sim2$ \citep[][]{hayashi09,lehnert09}) and low redshift
(grey circles: nearby sample of warm IR-luminous galaxies from
\citet{kewley01}. 
Contours show the density
distribution of emission line galaxies in the SDSS. Lines show the
boundary relations between star-forming galaxies and AGN
\citep[][dashed line]{kauffmann03} and composite galaxies and AGN
\citep[][solid line]{kewley06}, and the SDSS star-forming ridge line 
from \citet{brinchmann08a}. LBA 092159 (white cross) was found to
contain a compact radio core using VLBI (in prep.). See Sect. \ref{sec:agn} for details.}
\end{figure*}

\section{Discussion of the Emission Line Properties of LBAs}

The most significant results provided by our HST data are:\\

1) Highly luminous, compact star-forming ``clumps'' appear to be the
defining feature of the UV morphology of the LBAs.\\

2) In the most extreme cases, the UV images are dominated by a single
extremely bright and compact object surrounded by a disturbed envelope
or disk. We have called these dominant central objects (DCOs).\\

3) The star-formation rate derived from the extinction-corrected
H$\alpha$ emission-line luminosity tends to be systematically smaller
than that derived using the mid-IR and/or the extinction-corrected
far-UV continuum luminosities (see panel {\it f} of
Fig. \ref{fig:panels1} and Fig. \ref{fig:sfrs}). More properly, the
ratio of the H$\alpha$ luminosity to either the mid-IR or far-UV
continuum is smaller on-average than in typical star-forming
galaxies.\\

For our discussion below it is important to address first whether any of
these properties could be explained by the contribution from a Type 1
(unobscured) or Type 2 (obscured) AGN.

\subsection{Are they AGN?}
\label{sec:agn}

\subsubsection{Type 1 AGN}

Could the DCOs be Type 1 AGN in which we have a direct view of the
accretion disk and broad emission-line region? This would account for
the exceptionally compact morphology as well as for the significant
24$\mu$m ``excess''. We note that the initial LBA selection process
excluded any obvious Type I AGN in the first place based on the
absence of obvious broad (several thousand km s$^{-1}$) Balmer
emission-lines from the AGN Broad Line Region in the SDSS spectra
\citep{hoopes07}. We can quantify this and exclude the possibility
that some of our galaxies are of type Seyfert 1.8 or 1.9, with weak
broad emission lines. In the event that possible broad emission was
somehow being missed in the integrated SDSS spectrum, we have obtained
integral field spectra using FLAMES on the VLT. In
Fig. \ref{fig:haspec} we show the preliminary results of the
\ha$+$\nii\ line profile as measured in the central
$\sim$1\arcsec\ region centered on the clumps in four LBAs hosting
some of the most luminous clumps (three of which are DCOs). The line
profiles were fitted with a series of narrow and broad Gaussian
components indicated by red and blue dashed lines, respectively. The
fits show that each source has significant line asymmetries on the
blue side of the narrow lines that are consistent with a relatively
broad, blueshifted component (Overzier et al., in prep.). Such blue
asymmetries of width of a few hundred km s$^{-1}$ in both the
permitted and forbidden are not related to the much broader line
profiles from Type 1 AGN, which are seen only for the permitted
lines. Instead, the broad blueshifted wings seen on the emission-line
profiles in our objects (Fig. \ref{fig:haspec}) are signposts of
galactic winds and are commonly observed in intense local and high
redshift starbursts
\citep[e.g.][]{heckman90,lehnert96,lehnert09,shapiro09}.

\subsubsection{Type 2 AGN}

\begin{figure*}[t]
\begin{center}
\includegraphics[width=0.7\textwidth]{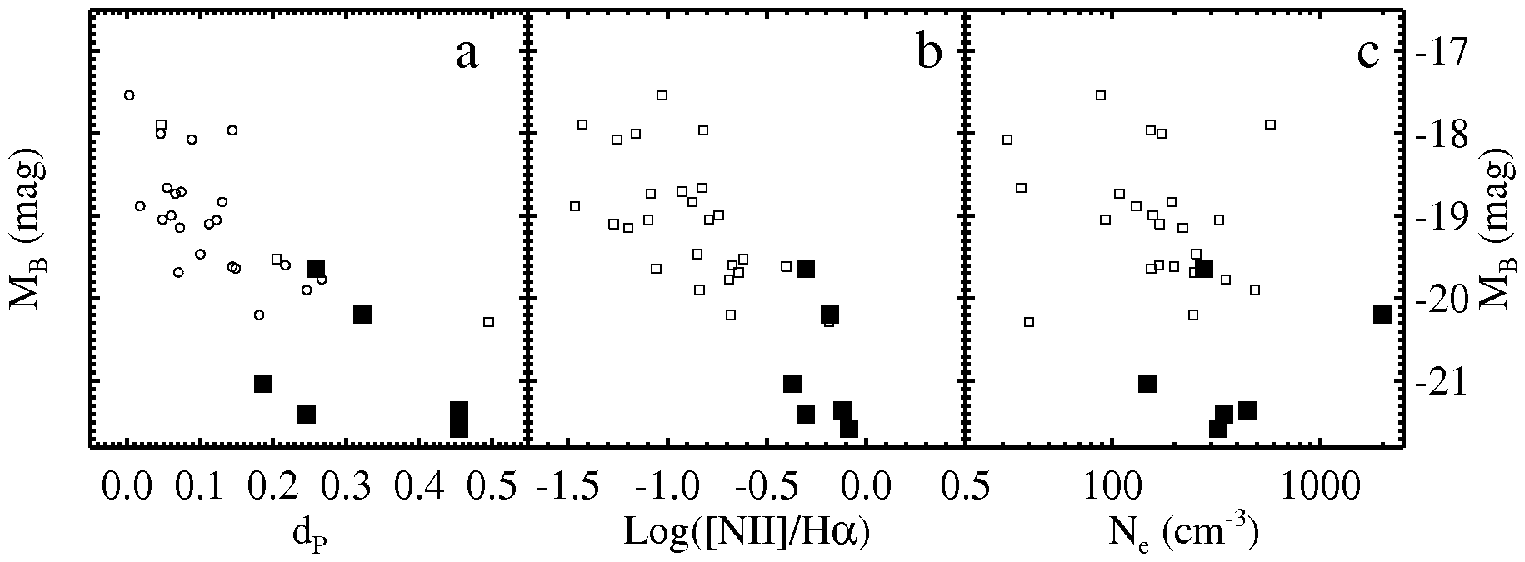}
\end{center}
\caption{\label{fig:panel2} Correlations between the absolute $B$-band
  magnitudes ($M_B$) of the star-forming clumps and various emission
  line properties of the host galaxies measured from the SDSS
  spectrum. In each plot, the six dominant central objects (DCOs) are
  indicated by filled squares. Panels show {\it (a)} $M_B$ versus BPT
  diagram offset $d_P$, {\it (b)} log(\nii/\ha), and {\it (c)} electron density $N_e$. The host
  galaxies of the brightest clumps tend to have larger BPT
  offsets $d_P$, larger log(\nii/\ha), and larger electron densities $N_e$. See
  \S\ref{sec:agn} for details.}
\end{figure*} 

Our spectra show clearly that the optical continuum must be dominated
by the light of young stars (with no detectable contribution from an
unobscured AGN). However, is it possible that the DCOs are associated
with the presence of an obscured (Type 2) AGN? In Fig. \ref{fig:bpt}
we show the log(\nii/\ha) versus the log(\oiii/\hb) diagnostic ``BPT''
diagram that is most commonly used for identifying the main source of
ionization among samples of star forming galaxies and AGN
\citep{baldwin81,kauffmann03,kewley06}. Because it is based on ratios
of bright emission lines that are observable out to $z\lesssim3$, the
diagram is widely used to characterize galaxies at both low and high
redshift. It has been known for some time that intensely star-forming
galaxies at high redshift such as Lyman breaks, ``s$BzK$'', and
distant red galaxies are often peculiarly offset towards higher line
ratios \citep[e.g.][]{teplitz00,erb06a,kriek07,lehnert09,hayashi09}
with respect to the mean SDSS star-forming population at low redshift
(Fig. \ref{fig:bpt}, dotted line).  Similar offets, albeit perhaps
slightly less extreme, have been observed in local ``warm'' IR-luminous
galaxies \citep{kewley01} and Wolf-Rayet galaxies
\citep{brinchmann08b}, and DEEP2 galaxies at $z\sim1.4$
\citep{shapley05b,liu08}. At present it is not known what physical
mechanisms lie at the root of these offsets \citep[see the detailed
  discussion in][]{brinchmann08a}, but it is important to note that
our sample (filled circles) shows similar offsets in the BPT diagram
(see also Hoopes et al. (2007) and Paper I). The BPT displacement is
in the sense of enhancements in one or both of the two line ratios.
If we focus on the locations of the host galaxies of the DCOs in
Fig. \ref{fig:bpt} (filled squares), we see that they are more
significantly displaced (in log[\nii/\ha]) with respect to the locus
of normal star-forming galaxies and the other LBAs. Typical galaxies
in the SDSS with similar emission-line ratios are almost certainly
composite objects in which the SDSS fiber encompasses both gas near
the nucleus that is excited by an AGN and normal regions of star
formation in the surrounding galaxy
\citep[e.g.][]{kauffmann03,kewley06,kauffmann09}. Could the DCOs
harbor a Type 2 AGN surrounded by a starburst?

To answer this question we further investigate the BPT offsets in
Fig. \ref{fig:panel2}. In panel {\it (a)} we plot for each clump the
dimensionless parameter $d_P$, defined as the relative offset its host
galaxy has in the BPT diagram. This offset distance $d_P$ is measured
perpendicular to the SDSS star-forming ``ridge'' (dotted line in
Fig. \ref{fig:bpt}).  This clearly demonstrates a relation between the
offsets and the luminosity of the clumps. Panel (b) shows the offsets
along log(\nii/\ha) only, but we note that the correlation here is driven
mainly by the well-known relation between mass and metallicity. 
In panel {\it (c)} we plot
$M_B$ versus the electron density ($N_e$) of the gas as derived from
the density sensitive \sii$\lambda\lambda$6713\AA,6731\AA\ line
doublet assuming a gas temperature of $10^4$ K \citep{osterbrock06}. The densities derived from the integrated SDSS
spectrum correlate with the $M_B$ of the clumps over almost two orders
of magnitude (the lowest and the highest being $\sim$30 and $\sim$2000
cm$^{-3}$) albeit with quite a large scatter. The implied gas
pressures are $P/k \sim 10^6$ to $10^7$ K cm$^{-3}$, several orders of
magnitude higher than in normal star forming galaxies in the local
universe.

\begin{figure*}[t]
\begin{center}
\includegraphics[width=0.3\textwidth]{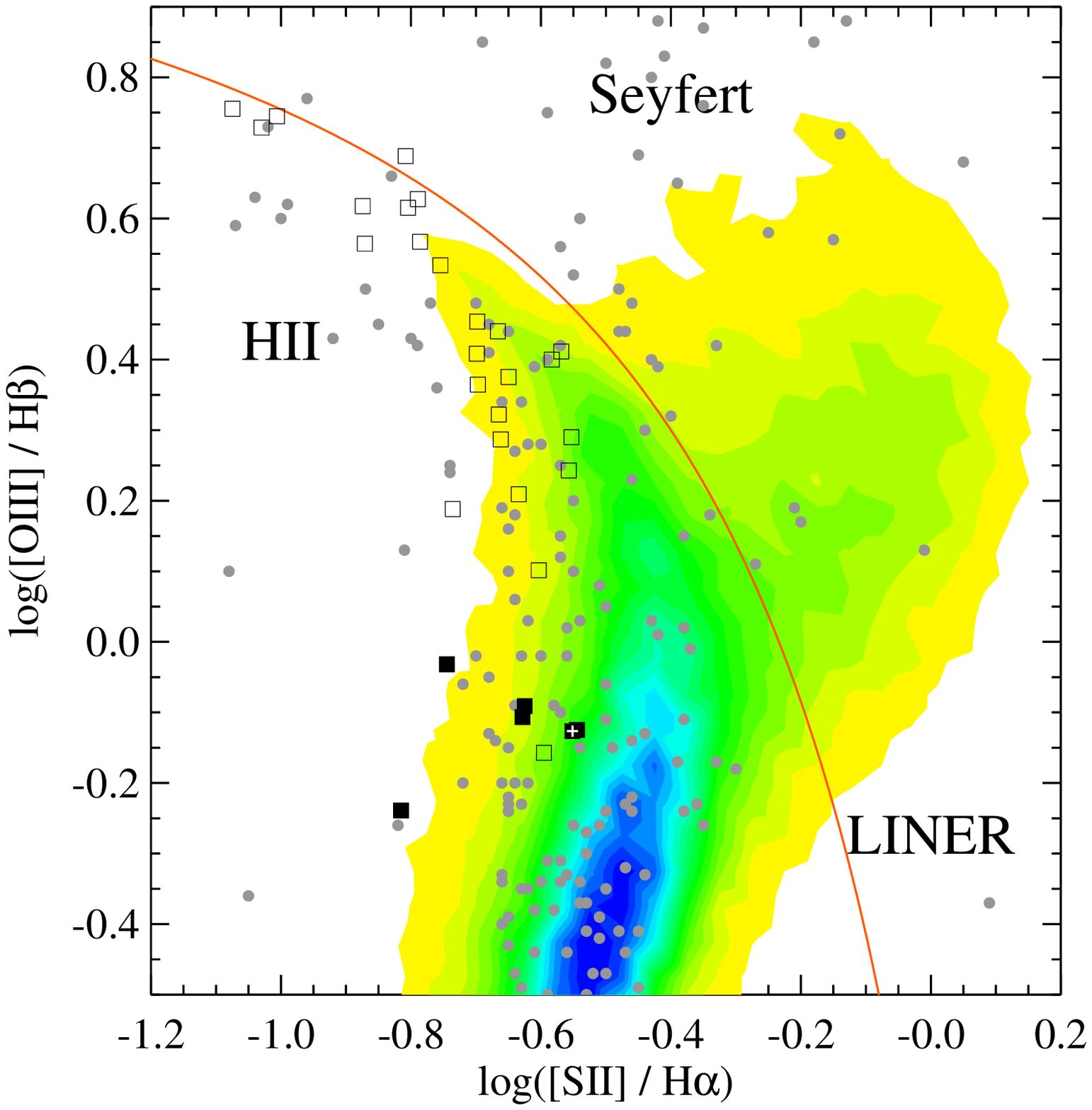}
\includegraphics[width=0.3\textwidth]{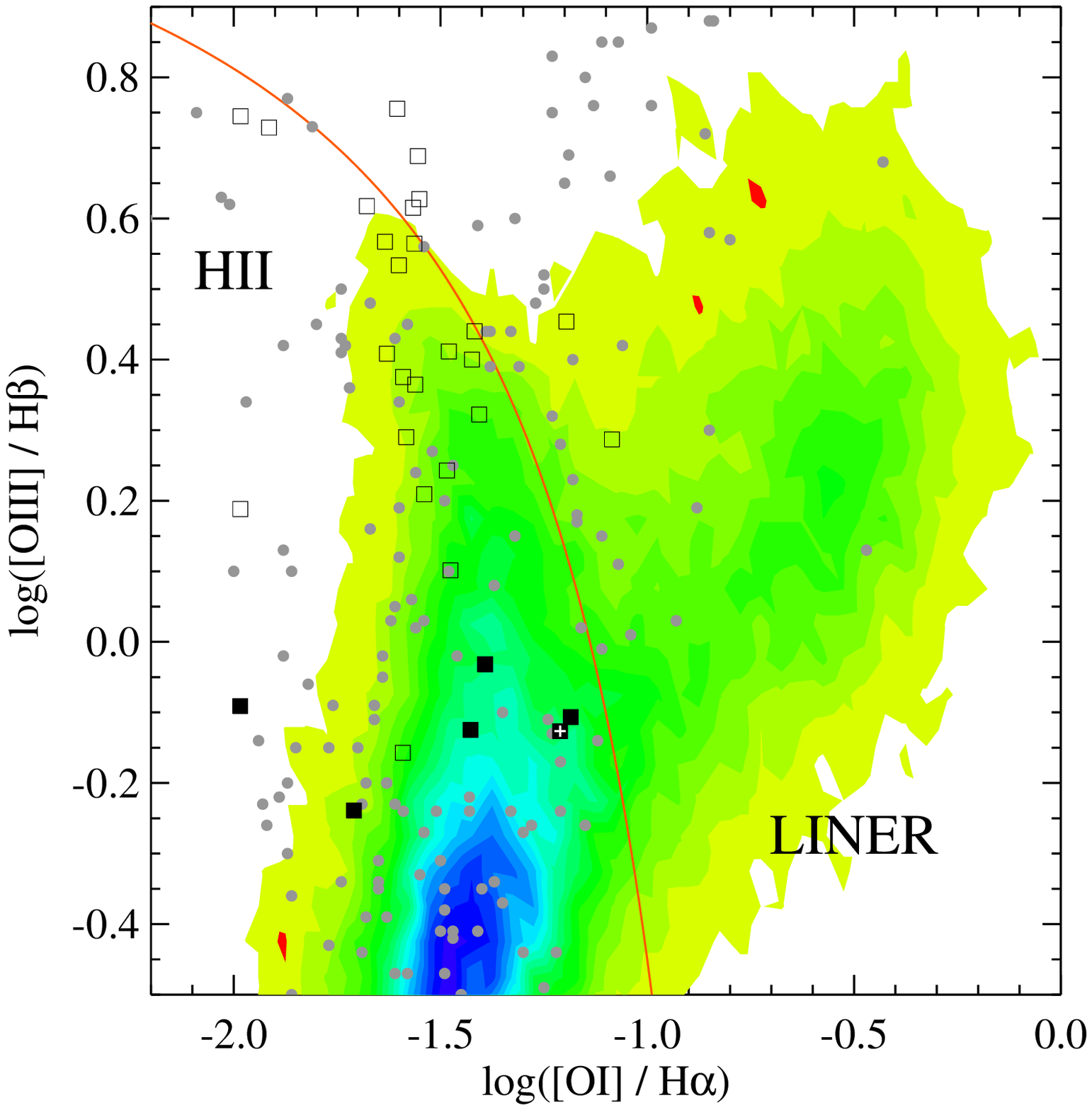}
\includegraphics[width=0.3\textwidth]{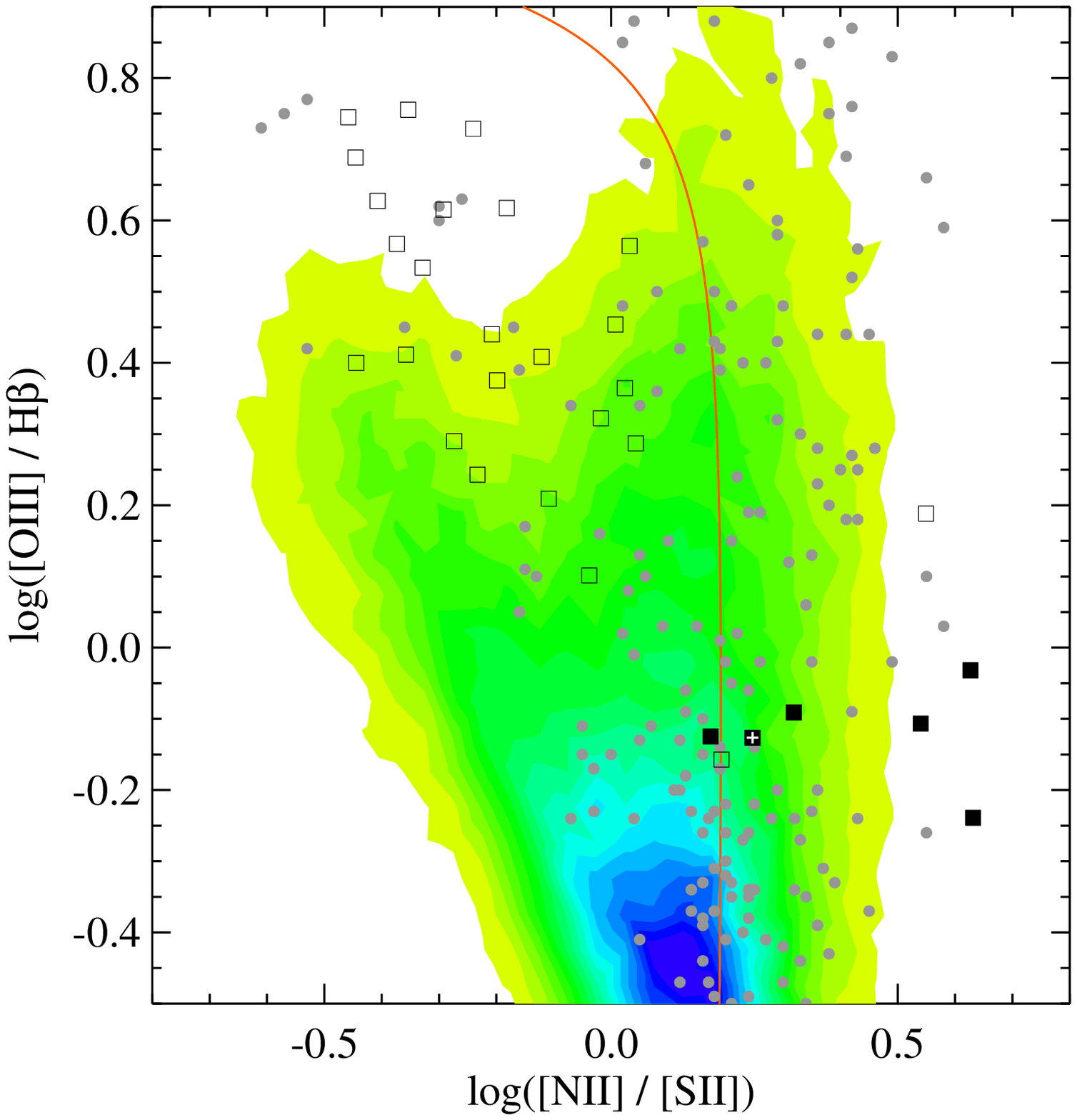}
\end{center}
\caption{\label{fig:bpt2}In order to further test the AGN hypothesis
  we show plots of \siib/H$\alpha$ vs. \oiiib/H$\beta$ and
  \oib/H$\alpha$ vs. \oiiib/H$\beta$
  \citep[following][]{veilleux87,kewley06}. Bona fide starburst-AGN
  composite systems will lie in the region intermediate between the
  locus of pure star-forming galaxies and pure AGN in these plots.
  Instead, we see some of the galaxies with the more massive DCOs lie
  in a kind of ``no man's land'' displaced from the locus of star
  forming galaxies in a direction opposite to that of the AGN (they
  have weaker \siib\ and \oib\ emission than normal star forming
  galaxies). Their anomalous behavior is even more dramatically
  illustrated in the right-most panel where we plot
  \niib/\siib\ vs. \oiiib/H$\beta$. Grey circles indicate the nearby
  warm IR-luminous sample from \citet{kewley01}. See the text in
  Sect. \ref{sec:agn} for details.}
\end{figure*}
 
The Type 2 AGN interpretation would be consistent with the high
electron densities seen in the most extreme cases ($\sim10^3$
cm$^{-3}$), since such densities are typical in the narrow
emission-line region in AGN \citep{osterbrock06}.  It would also be
consistent with the blue-asymmetric emission-line profiles seen in
Fig. \ref{fig:haspec}, which are characteristic of outflows in emission-line AGN
\citep[e.g.][]{heckman81,whittle88}. The AGN hypothesis could also
explain why the ratio of the 24 micron and extinction-corrected
H$\alpha$ luminosity is higher than in typical star forming galaxies
(Fig. \ref{fig:sfrs}), since AGN are characterized by strong emission
by warm dust peaking in the mid-IR \citep[e.g.][]{spinoglio02}. It
would not however explain the result in Fig. \ref{fig:sfrs} that the
ratio of the extinction-corrected far-UV and H$\alpha$ luminosity is
also higher on-average in the galaxies with DCOs than in typical
star-forming galaxies, since in a Type 2 AGN the continuum from the
AGN itself will be very heavily obscured in the far-UV.

To further test this starburst/AGN composite hypothesis, we can use
other emission-line diagnostic diagrams that are complementary to the
BPT diagram. Following \citet{veilleux87} and \citet{kewley06} we show
in Figure \ref{fig:bpt2} plots of \sii/H$\alpha$ vs. \oiii/H$\beta$
and \oi/H$\alpha$ vs. \oiii/H$\beta$. Bona fide starburst-AGN
composite systems will lie in the region intermediate between the
locus of pure star-forming galaxies and pure AGN in these plots.
Instead, we see that the galaxies with the DCOs tend to lie in a kind
of ``no man's land''  displaced from the locus of star forming
galaxies in a direction opposite to that of the AGN (they have weaker
\sii\ and \oi\ emission than normal star forming galaxies). Their
anomalous behavior is even more dramatically illustrated  in the
right-most panel in Fig. \ref{fig:bpt2} where we plot
\nii/\sii\ vs. \oiii/H$\beta$. As indicated in
Figs. \ref{fig:bpt} and \ref{fig:bpt2} (small grey circles), 
similar behavior is seen in local IR-warm starbursts
\citep{kewley01}.

Thus, the hypothesis that the compact clumps are found in
starburst/AGN composite systems does not explain the properties of
these objects. In future papers we will describe observations with
XMM-Newton, the Spitzer Infrared Spectrograph, and the European VLBI
network that are designed to search for more direct signatures of the
presence of an AGN.

\subsection{Further Clues from the Emission Line Properties}
\label{sec:lines}

We have estimated the ages of the clumps and DCOs to be $\sim$5--100
Myr assuming a model of an instantaneous burst of star formation (see
Table \ref{tab:clumpprops}). We have also shown that there exist
striking correlations between the properties of the clumps (e.g.,
luminosity), the properties of their host galaxies (e.g., mass,
radius, and concentration) and the physical properties of the
surrounding gas (e.g. ionization and density). The LBAs with DCOs are
particularly extreme in these plots. Below we will discuss the most
likely physical mechanisms that can explain these correlations.

The results presented in Section 3 showed that there are significant
correlations between the properties of the massive compact clumps and
the physical properties of the surrounding ionized gas as traced by
the nebular emission-lines. In particular, the DCOs tend to be found
in galaxies in which:\\

1. The star-formation rate derived from the extinction-corrected
H$\alpha$ emission-line luminosity tends to be systematically smaller
than that derived using the mid-IR and/or the extinction-corrected
far-UV continuum luminosities (see panel {\it (f)} in
Fig. \ref{fig:panels1} and Fig. \ref{fig:sfrs}). More properly, the
ratio of the H$\alpha$ luminosity to either the mid-IR or far-UV
continuum is smaller on-average than in typical star-forming
galaxies.\\

2. The location of the galaxy in the diagnostic plot of
\oiii/H$\beta$ vs. \nii/H$\alpha$ (BPT diagram, Fig. \ref{fig:bpt})
is more significantly displaced with respect to the locus of normal
star-forming galaxies (see panels {\it (a)} and {\it (b)} in
Fig. \ref{fig:panel2}). This displacement is in the sense of
enhancements in one or both of the two above ratios.\\

3. The electron density in the ionized gas is significantly higher
than in the other galaxies (panel {\it (c)} in
Fig. \ref{fig:panel2}).\\

Occam's razor drives us to ask if there is a single explanation for
all of these results. In Section \ref{sec:agn} we have already argued
that the contribution from an obscured, low-luminosity AGN could
explain some but not all of the properties observed. Here we will
consider two alternative possibilities: 1) The DCOs are short-duration
starbursts caught at a time after the peak in the O-star population,
and 2) The DCOs include ``leaky'' systems in which a large fraction of the
Lyman continuum is escaping.

\subsubsection{The Aged Starburst Hypothesis}

The DCOs are extremely massive ($\sim10^9$ M$_{\odot}$), compact
($\sim 10^2$ pc) objects. These high densities ($\sim10^3$ M$_{\odot}$
pc$^{-3}$) imply very short dynamical times ($R/v \sim$ 1 Myr). On
simple grounds of causality, it is therefore plausible that these
objects could have formed all their stars in a Myr, a timescale
significantly less than the characteristic lifetimes of the O stars
that produce nearly all the ionizing radiation in a starburst. In such
a case, the ratio of the ionizing luminosity to the non-ionizing
far-UV luminosity changes rapidly with time, and falls dramatically
for ages greater than about 5--6 Myr \citep[e.g.][]{leitherer95}. The
de-reddened UV/optical colors of the massive compact clumps shown in
Fig. \ref{fig:cc} (while uncertain) are consistent with those of such
``aged'' instantaneous starbursts with ages of one-to-a-few
tens-of-Myr (see Table \ref{tab:clumpprops}). This is consistent with
\citet{basu-zych07} who compared the SFRs derived from the UV, IR and radio. We should note however
that at least in one case (210358) we find evidence for WR stars in the SDSS
spectrum indicating that its SFH must include a much more recent ($<5$
Myr) burst.

If, however, the aged starburst hypothesis is correct for other
objects, this would naturally explain the relative weakness of the
H$\alpha$ emission-line in these galaxies. Such a model could also
explain the peculiar location of these objects in the BPT diagram
(Fig. \ref{fig:bpt}). For an instantaneous burst in the age range
between about ten and fifty Myr, the primary heating/ionizing source
for the associated gas would be the mechanical energy supplied by
supernovae rather than O star ionizing radiation \citep{leitherer95}.
Models of such shock-heated gas \citep[e.g.][]{dopita95,allen08} show an
enhancement in the \oiii/H$\beta$ and \nii/H$\alpha$ ratios, as we
observe (Fig. \ref{fig:bpt}).

The expected supernova rate would be $\sim4$ ($M_{clump}/10^9$
M$_\odot$) per decade. Such a high supernovae rate occurring in a very
small region also naturally accounts for the high electron densities
and thus the high gas pressures ($P/k \sim 10^7$ K cm$^{-3}$) in these
objects. Following \citet{chevalier85}, in the absence of significant
radiative losses, the central pressure inside a starburst region of
radius $r_{100}$ (in units of 100 pc) with a supernova rate of
$R_{SNe}$ per decade is given by $P/k = 1.4\times10^8 R_{SNe}
r_{100}^{-2}$ K cm$^{-3}$. \citet{chevalier85} also demonstrate that
this high-pressure gas will subsequently expand outward to form a
galactic wind with a ram pressure that drops like $r^{-2}$. This wind
would be sufficient to produce the observed gas pressures over
kpc-scales in the region surrounding the clump. The galactic wind will
also accelerate emission-line clouds, leading to the broad
blue-shifted wings seen on the emission-line profiles in our objects
(Fig. \ref{fig:haspec}). In fact, similar high gas pressures, broad
blueshifted wings, and other signposts of galactic winds are common in
intense local and high redshift starbursts
\citep[e.g.][]{lehnert96,lehnert09,shapiro09}.

This model fails however to explain the relative weakness of the
\sii\ and \oi\ lines seen in Fig. \ref{fig:bpt2}. Shock heating should
lead to an enhancement in the relative strengths of these lines
\citep[see Fig. 2 in][]{dopita95}. We are left with an otherwise
attractive model that does not naturally explain all the properties of
these objects.
 
\subsubsection{The Leaky Starburst Hypothesis}

An obvious way to lower the amount of observed H$\alpha$ emission
would be a ``leaky'' interstellar medium in which a significant
fraction of the ionizing radiation produced by O stars escapes the
galaxy. This hypothesis has a plausible physical basis: the feedback
effect of concentrating the release of the kinetic energy from
of-order $10^7$ supernovae ($\sim10^{58}$ ergs) all within the
exceptionally small volume of a DCO may create large holes in the
surrounding interstellar medium through which ionizing radiation could
then escape. Taken at face value, Fig. \ref{fig:sfrs} would then imply
that the majority of the ionizing radiation is escaping in our two
most extreme objects (021348 and 080844). It is interesting to note
that \citet{shapley06} find 2 out of 14 LBGs at $z\sim3$ with a
significant escape fraction ($f_{esc}\simeq50-100$\%) as measured from
the ratio of the fluxes below and above the Lyman limit in UV
spectra. Similarly, \citet{iwata09} report the detection of large
$f_{esc}$ in 7 of 73 LBGs at $z\sim3$. These fractions are similar to
what is implied for our sample of LBAs under the leaky starburst
hypothesis.

The leaky starburst hypothesis can readily account for the high
electron densities (gas pressures) and the broad blue wings on the
emission-line profiles: these are still intense starbursts and the
same arguments as given above will apply. A leaky interstellar medium
could also explain at least some aspects of the peculiar emission-line
ratios we observe. Models of photoionized gas under ``matter-bounded''
conditions
\citep[e.g.][]{binette96,castellanos02,giammanco05,brinchmann08a} show
that the strengths of the \oi\ and \sii\ lines relative to H$\alpha$
are decreased. This is because these lines are relatively strong in
the warm partially neutral region lying beyond the edge of the
Stromgren sphere, and this region is missing in matter-bounded
conditions. These models show that the \oiii/H$\beta$ ratio will also
be enhanced in matter-bounded clouds, as we observe (Fig.
\ref{fig:bpt}). However, this model fails to explain the enhanced
\nii/H$\alpha$ ratios seen for the DCOs in the BPT diagram. Models (B. Groves,
private communication) show that matter-bounded clouds in starbursts
should have \nii/H$\alpha$ ratios that are similar to or lower than
ionization-bounded clouds.

A more fundamental objection to the leaky starburst hypothesis is that
is difficult to envisage a situation in which a sizable fraction of
the ionizing radiation escapes the galaxy, but the majority of the
non-ionizing UV radiation is absorbed by dust inside the galaxy (as is
required by the high mid-IR luminosities we see). For this reason, we
regard the ``aged starburst'' hypothesis as the more likely one.

In the future, we will use longslit and integral field spectroscopy in order to 
spatially resolve and disentangle the line emission coming
from the clumpy versus the diffuse regions seen in the HST images.  

\section{Discussion of the Structural Properties of LBAs}
\label{sec:clumps}

The peculiar {\it emission line properties} discussed above show the
presence of significant feedback processes between the massive
star-forming regions in LBAs and their surrounding ISM. Here, we will
focus on the remarkable {\it stellar properties} of the massive clumps
and the DCOs in particular, by asking the question to what kind of
structural components of galaxies do they correspond? 

The DCOs are only marginally resolved in the HST UV images, with
implied radii of no greater than $\sim$100 pc (see
Fig. \ref{fig:psf}). The fainter clumps are typically several times
larger than this. The masses of the DCOs estimated in Section 3 range from a few
$\times 10^8$ to a few $\times10^9$ M$_{\odot}$, while the more
typical clumps have masses in the range of a few $\times10^7$ to a few
$\times 10^8$ M$_{\odot}$.  This implies very high stellar mass surface
densities. For the typical clumps values for the effective surface
mass density ($\Sigma_e\approx M_{*,clump}/(2\pi R_e^2)$) range from
$\sim10^8$ to $10^9$ $M_\odot$ kpc$^{-2}$, while for the DCOs
$\Sigma_e \sim 10^{9.5}$ to $10^{11}$ $M_\odot$ kpc$^{-2}$. What then
will the clumps and DCOs evolve into? Can we identify such evolved
counterparts in local galaxies?

\subsection{Super Star Clusters and Central Massive Objects?} 

Massive clusters are observed in and around the nuclei of many nearby
late- and early-type galaxies, mergers and merger remnants, and AGN
\citep[e.g.][]{carollo97,geha02,boker04,maraston04,walcher05,seth08}.
The largest stellar clusters typically have sizes of only a few pc
(the largest being $\approx$10 pc) and masses rarely exceeding
$\simeq10^7$ $M_\odot$. Their structural properties are thus similar
to those of massive globular clusters in the Milky Way and local
galaxies, and also to the young ``super star clusters'' (SSCs) found
in starburst galaxies \citep{meurer95}.  

The most luminous star clusters in normal and starburst galaxies lie
along a relation between the absolute optical magnitude of the cluster
and the global SFR of the host galaxy \citep{larsen02}, a relation
that is mainly driven by the relation between the luminosity of the
brightest cluster and the total number of clusters formed in a galaxy
\citep{whitmore07}.  The correlation implies that for galaxies having
SFRs as high as LBAs even the brightest clusters should only have
$M_V\simeq-16$ to --14
\citep[$M_V=-1.87\mathrm{log}(\mathrm{SFR})-12.14$, see][]{weidner04}
compared to $-21.5\lesssim M_V\lesssim-18$ found in Fig. \ref{fig:cc}. 
This discrepancy could be partly reconciled given that the LBA clusters are
systematically younger than the brightest clusters observed in normal
galaxies, since a star cluster fades by $\sim3$ mag in the first few 100
Myr. However, this scenario still largely falls short for any of the luminous
clumps/DCOs having $-21.5<M_V<-19$.

The most massive cluster-like object known to date, object W3 in the
merger remnant NGC 7252, has a mass of $(8\pm2)10^7$ $M_\odot$ and a
radius of $17.5\pm1.8$ pc \citep{maraston04}. Interestingly, its high
mass and small size are very much at odds with any kind of stellar
cluster found to date. Curiously, the properties of object W3 are
perhaps more similar to that of ultra-compact dwarf galaxies
\citep[e.g.][]{mieske04}. If we extrapolate from its current
luminosity at an age of $\sim$400 Myr to that expected at 10 Myr, we
find $M_V^{10Myr}\simeq-19.2$ \citep{bastian08}, still over 2
magnitudes fainter than some of the DCOs.

Although the stellar mass surface densities that we derive ($\sim10^{8.5}-10^{11}$
$M_\odot$ kpc$^{-2}$) are comparable to that of typical massive
clusters mentioned above, the sizes are about an order of magnitude larger
($\sim$100 pc) and the masses are up to two orders of magnitude larger
($\lesssim10^9$ $M_\odot$). Thus, the massive clumps and DCOs
identified in the LBA sample appear as significantly scaled-up
versions of known types of massive star clusters, and are not simply
an extension of the cluster luminosity function as derived from nearby
star-forming galaxies.

Furthermore, nuclear super star clusters and supermassive black
holes are jointly referred to as central massive objects (CMOs). CMOs
follow a tight correlation with the total mass of the galaxy over many
orders of magnitude in stellar mass. The central star clusters in low
luminosity early-types and spiral bulges, as well as the masses of the
central supermassive blackholes in $L\ge L_*$ early-types all lie on
the relation $M_{CMO}\approx0.002M_{gal}$ with small scatter
\citep[e.g.][]{magorrian98,marconi03,cote06,ferrarese06,wehner06,rossa06}. Simulations
of both isolated and merging galaxies reproduce the observed
correlation well \citep[e.g.][]{li07}. As we discuss below in more
detail, the clumps and DCOs typically represent several percent of the
total stellar mass of the surrounding galaxy. These mass fractions are
therefore over an order of magnitude higher than in the CMOs.

Could the massive clumps or DCOs consist of multiple SSCs? If so, how
long would it take this ensemble to merge into a single object? To
address this, we assume that a clump of mass $10^8$ $M_\odot$ actually
consists of 10--100 SSCs each having a mass of $10^{6-7}$ $M_\odot$
distributed over a region with a 100 pc radius within which the
velocity dispersion
$\sigma_{clump}\approx\sqrt{GM_{clump}/5R_e}\approx30$ km
s$^{-1}$. Then the timescale over which a single SSC migrates inward
as a result of dynamical friction is on the order of only $\sim$10--20
Myr \citep{binney08}, comparable to the typical ages we have derived
for the clumps. This suggests that a given massive clump (or DCO) is
likely to have evolved into a single very massive object, even if it
started out as an ensemble of SSCs.

We conclude that the clumps and DCOs are very different structurally
from typical nuclear (super) star clusters. 

\begin{figure*}[t]
\begin{center}
\includegraphics[width=0.7\textwidth]{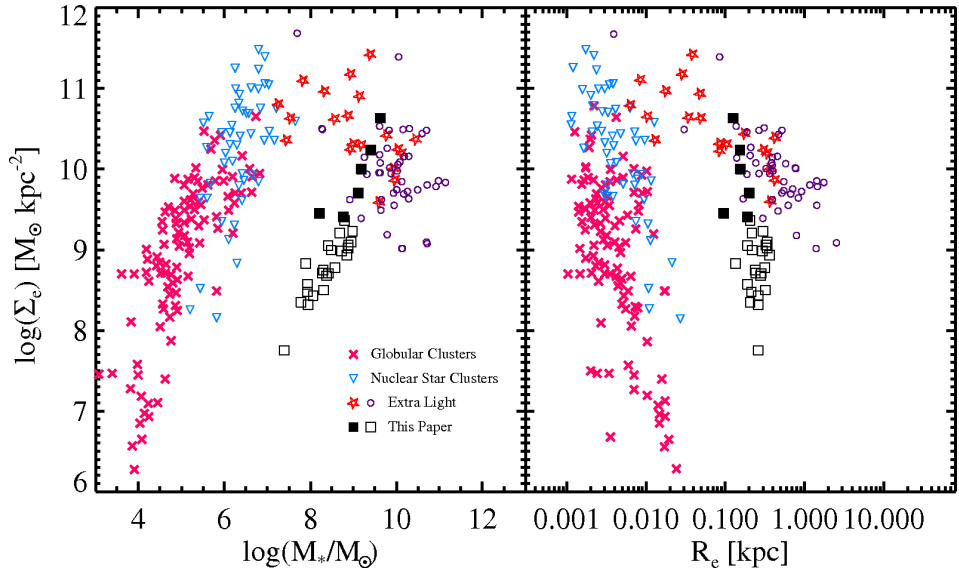}
\end{center}
\caption{\label{fig:hopkins}Mass ($M_*$) and effective radius ($R_e$)
  versus the effective mass surface density ($\Sigma_e$) of the
  brightest clump identified in each galaxy (squares). The comparison
  data shown was extracted from a compilation of literature results
  presented in Fig. 45 in \citet{hopkins09}, showing globular clusters
  (crosses), nuclear star clusters (triangles), and the extra-light
  components in early-type galaxies from the samples of
  \citet[][stars]{kormendy08} and \citet[][circles]{lauer07} as
  measured by \citet{hopkins09}. Objects that we identify as DCOs
  based on their location in the $M_*$, $R_e$, $\Sigma_e$ plane in
  combination with the large total stellar masses and concentrations
  of their host galaxies are indicated with filled squares. We suggest
  that DCOs could be the progenitor starbursts of the central ``extra
  light'' component in local cuspy elliptical galaxies. See
  Sect. \ref{sec:cusps} for details.}
\end{figure*}

\subsection{Extra light and central ``cusps'' in elliptical progenitors?}
\label{sec:cusps}

For the discussion below it is important to point out that the DCOs
(and, more generally, the most luminous clumps) tend to appear in the
more massive galaxies in our LBA sample (see panel {\it c} in
Fig. \ref{fig:panels1}). The DCOs also each sit in the middle of
extended, perturbed disk-like structures (hence the correlation
between clump luminosity and the optical concentration index, panel
{\it a} of Fig. \ref{fig:panels1}). Another characteristic (by
definition) is that the DCOs are single objects rather than the
multiple (fainter) clumps as seen in many of the other objects.

In Fig. \ref{fig:hopkins} we plot the clump mass ($M_*$) and effective
radius ($R_e$) versus the effective mass density ($\Sigma_e$) for the
brightest clump in each LBA, and compare with a collection of results
from the literature. The comparison data shown was extracted from a
compilation of literature results presented in Fig. 45 in
\citet{hopkins09}, showing globular clusters (crosses), nuclear star
clusters (triangles), and the extra-light components in early-type
galaxies from the samples of \citet[][stars]{kormendy08} and
\citet[][circles]{lauer07}. The high masses of $\sim10^9$ $M_\odot$
derived for the DCOs combined with the strong constraints on their
small effective radii ($\sim100$ pc as evidenced by the HST
diffraction spikes  in
Fig. \ref{fig:points}) yield the highest mass surface densities found
in our sample, $\Sigma_M\approx10^{10}-10^{11}$ $M_\odot$ kpc$^{-2}$
(see Table \ref{tab:massiveclumpprops}).

\citet{hopkins09} show that this region of parameter space in the
$M_*$ and $R_e$ versus $\Sigma_e$ planes is almost exclusively
occupied by the so-called central excess or ``extra light'' component
seen in local elliptical galaxies \citep[stars and circles in
  Fig. \ref{fig:hopkins};][and references
  therein]{crane93,lauer95,kormendy08}. This extra light is
characterized as an inner ($\lesssim$1 kpc) component that rises more
steeply than expected from the extrapolated Sersic power-law that fits
the outer profile of ellipticals well. The so-called ``cusps'' have
effective radii of $\sim50-500$ pc and show a trend of decreasing
density with increasing mass and in that sense they lie on the same
sequence formed by elliptical galaxies. The cuspy cores are
furthermore typical for relatively low-mass ($\le L_*$) ellipticals,
while they are relatively uncommon for the most massive
ellipticals. \citet{kormendy08} suggest that the core--cusp dichotomy
in local ellipticals may correspond to their formation history: the
extra light or cusp components are very likely the remnant signature
of a massive central starburst in a dissipative (``wet'') merger,
while the extended cores in core-dominated ellipticals are consistent
with being the relaxed descendants of dissipationless (``dry'')
mergers.

While none of the clumps bear any structural similarities with both
globular and nuclear star clusters (crosses and triangles), the DCOs
in the $M_*$, $R_e$, $\Sigma_e$ plane appear structurally similar to
the extra-light light components in early-type galaxies (stars and
circles).
\citet{hopkins09} investigate in detail how the central mass
associated with the extra light component is related to the remnant of
a massive central starburst fueled by dissipative merging, and use
extensive simulations to show that such a mechanism successfully
reproduces the observed profiles of elliptical galaxies
\citep{hopkins08,hopkins09}. In particular, they find a strong
correlation between the galaxy stellar mass and the fraction of the
total stellar mass that is in the extra light component ($f_{extra}$),
as well as a one-to-one correlation between the extra light mass
fraction and the fraction of the total stellar mass that formed in the
dissipative starburst ($f_{starburst}$), with roughly a factor of $\sim2$ intrinsic scatter in $f_{starburst}$
and a factor of $\sim4$ scatter in $f_{extra}$ at each $M_*$.  
For stellar masses in the range of $10^{10}$
to $10^{11}$ $M_\odot$ the $f_{starburst}(f_{extra})$ ranges from a
few to 50\%, consistent with the extra-light fractions measured in
real ellipticals by \citet{hopkins09}. 

\begin{figure}[t]
\begin{center}
\includegraphics[width=\columnwidth]{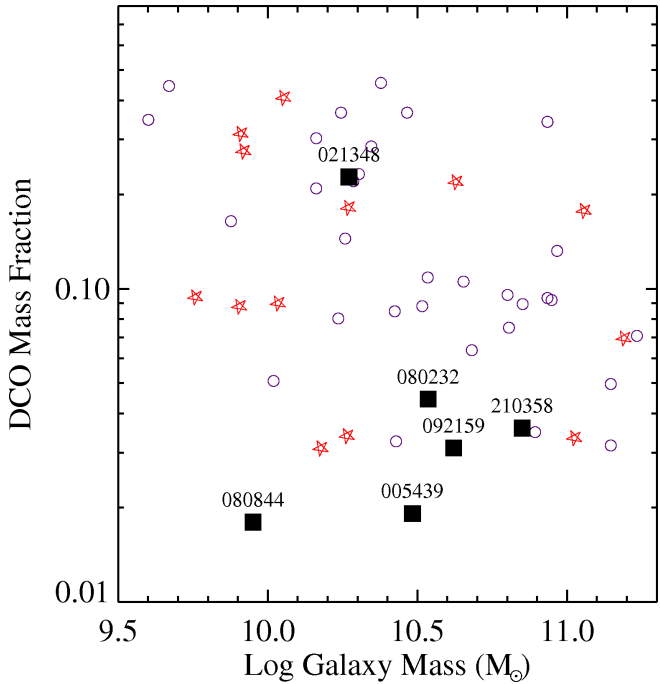}
\end{center}
\caption{\label{fig:dco}Fraction of the stellar mass in the DCOs (see
  Fig. \ref{fig:points}) 
  versus the total stellar mass of the six objects. The DCOs contain 2--23\% of the total
  stellar mass, suggesting that they could be the hypothesized
  progenitor starbursts related to the central ``extra light''
  component observed in local cuspy elliptical galaxies. This is also
  consistent with their total stellar masses of $10^{10}\lesssim
  M_*/M_\odot\lesssim10^{11}$. The comparison data shown (stars and
  circles, taken from Fig. 20 in \citet{hopkins09}) correspond to the
  same extra-light components as plotted in
  Fig. \ref{fig:hopkins}. See Sect. \ref{sec:cusps} for details.}
\end{figure}

We can compare this with the ratio of the central clump mass
to the total stellar mass we found for the DCOs. Due to the uncertainties in the mass estimates
involved, we estimate the DCO mass by taking the average of the values
derived using single burst and continuous SFHs.
The result is shown in Fig. \ref{fig:dco}. We find mass fractions
typically of a few percent ($f_{DCO}\sim2-4$\%), object
021348 being a notable exception ($f_{DCO}\sim23$\%, see also Table
\ref{tab:massiveclumpprops}). Except for object 021348, which agrees
well with the cuspy cores, the mass fraction of the DCOs appears to be relatively low compared to
the early-type cusps plotted in Fig. \ref{fig:dco} (stars and circles). However, the mass
fractions we find are still significant and not inconsistent with the data. 

Alternatively, we should note that the special region of parameter
space in $M_*$, $R_e$ and $\Sigma_e$ occupied by both the DCOs and the
cuspy cores appears to represent a particular transition region
between massive, 
cluster-sized objects on one hand and entire galaxies on the other. While low-mass star clusters
follow the scaling relations of globular clusters for masses
$\lesssim10^6$ $M_\odot$, a wide variety of higher mass objects
ranging from the most massive nuclear star clusters, dwarf-galaxy
transition objects, ultra-compact dwarf galaxies, dwarf elliptical
nuclei, and the early-type cusps themselves all appear to lie on the same
scaling relations for spheroidal galaxies
\citep[e.g.][]{kissler06,hopkins09}. Although this overlap may be partially
coincidental, perhaps it can be explained if these objects are all the
product of some generic dissipational process.

In contrast with the DCOs, the much less extreme clumps indicated in Fig. \ref{fig:hopkins}
are typically found in LBAs containing multiple clumps distributed
within their central regions (see Figs. \ref{fig:stamps} and 
\ref{fig:clumps}). For a clump of mass
$M_{clump}\sim10^8$ $M_\odot$ moving within a disk of mass
$M_{disk}\sim10^{10}$ $M_\odot$ and a disk radius of $R_{disk}\sim$1.5
kpc (the average optical half-light radius) we find
$\sigma_{disk}\approx\sqrt{GM_{disk}/3.4R_{disk}}\approx80$ km
s$^{-1}$ and a dynamical friction time of $\sim$100--200 Myr. This is
several times longer than the ages that we have derived for the
individual clumps. Thus it is plausible that at this relatively early
dynamical stage we usually see several such clumps within the
effective radius of the galaxy. The diffuse light seen in the HST
images may build up as the clumps lose stars due to tidal stripping
while they interact with each other and with the disk and slowly
migrate inward. This process may eventually lead to the formation of
an inner bulge as well as the Sersic profile typical of disk galaxies
\citep{noguchi99,immeli04,carollo97,bournaud07,bournaud08,elmegreen08a,elmegreen08b,elmegreen09}.

\subsection{Relation to Mergers and IR-Luminous Galaxies}

Although most LBAs were unresolved in the original SDSS images, some
showed evidence for close companions or faint extended emission.  One
of the major results from the HST images is that most of them 
show very complex morphologies at scales well below the SDSS seeing or
sensitivity.  We interpret the large array of morphologies in
Fig. \ref{fig:stamps} as evidence that most LBAs are undergoing or
have undergone an interaction or merger\footnote{Here we use the term
  ``merger'' in the widest sense of the word, i.e. it is not limited
  to ``major mergers'' but includes minor mergers and other mechanisms
  that may deliver a supply of gas-rich material.}. 
We also see a small number of objects that do not show unambiguous
evidence of merging such as tidal tails or companions, but that are
nonetheless often highly peculiar as evidenced by ring or chain-like
morphologies (e.g. object 102613 and objects 082001, 082550, 083803).
These observations suggest that the starbursts in LBAs and the
formation of the massive stellar clumps are somehow linked to the
mergers \citep[e.g.][]{larson78,lambas03,nikolic04,woods06,owers07,li08}. The
LBAs must also be sufficiently gas-rich in order for any starburst
trigger to be effective. Simulations show that the inflow of gas
ensures that the pressure in the interstellar medium becomes larger
than the internal pressure of giant molecular clouds, and that the
clouds can collapse and form massive clusters before they are disrupted by supernova explosions
\citep{bekki04}.

It is important to note that the spectral energy distributions of LBAs
are very different from typical starburst galaxies in the local
Universe. These are predominantly associated with the highly dust-obscured LIRGs ($10^{11} \le L_{IR}
\le 10^{12}$ $L_\odot$) and ULIRGs ($L_{IR} > 10^{12}$ $L_\odot$). 
At least ULIRGs are known to be strongly associated with the different phases
of galaxy--galaxy mergers, with about equal fractions showing dual
nuclei (pre-merger) and showing indication of post-merging (after the
separate nuclei have coalesced and a period of violent relaxation
starts forming an extended spheroidal component)
\citep[e.g.][]{genzel01,tacconi02,dasyra06a,dasyra06b,arribas08}. These
mergers are found to be limited to galaxy pairs having near-equal mass
ratios (1.5:1 on average, with a maximum of 3:1). The less luminous
LIRGs are typically associated with much less violent events that are
less effective at triggering a major starburst phase as observed in
ULIRGs, perhaps as a result of smaller gas fractions, orbital
parameters, and larger pair mass ratios (or a combination thereof). As
shown in \citet{arribas08}, about one third of LIRGs appear in single
isolated (spiral) galaxies.
While none of the LBAs qualify as ULIRGs, there is some overlap with
the LIRGs since $\sim$30--50\% of
LBAs have $L_{IR}>10^{11}$ $L_\odot$. As such, LBAs likely represent a
population of relatively unobscured mergers or merger remnants that
share some characteristics with LIRGs. The number density of LBAs is
much lower than that of LIRGs. If the two populations are somehow
connected, the LBAs could represent a very short evolutionary phase
during which the starburst can be seen relatively
unobscured. Alternatively, or in addition, LBAs might be seen at a
special viewing angle, perhaps along the axis where the starburst has
blown a hole in the ISM and is leaking Lyman continuum photons.      

\subsection{Implications for Black Hole Formation}
\label{sec:smbhs}

We have shown that the LBAs are consistent with galaxies that are in
the very early stages of forming their central bulge and/or core
excess. Local galaxies have tight correlations between the mass of
their central black holes and that of their bulge and total mass. It
is thus interesting to ask whether there is any evidence that a large
black hole mass is accumulating in the LBAs as well. The DCOs are
particularly intriguing in this regard: such very massive and compact
objects appear to be the ideal breeding gounds for a supermassive
black hole. As we have no means of directly detecting the
gravitational signature of a black hole inside the massive clumps, the
only way to infer the presence of a massive black hole would be
through the spectral signature of an actively accreting AGN.  As
discussed in Sect. \ref{sec:agn}, we did not find any direct evidence
for energetically significant AGNs in the DCOs.
 
It has long been suggested that the mass loss from stars and
supernovae in a very massive compact cluster at the center of a galaxy
may play a major role in the fueling of a supermassive black holes
\citep[][]{norman88}. While the stellar winds from O and WR stars and
the ejecta from core-collapse supernovae may in principle provide a
significant amount of material available for accretion during the
early stage of a massive central starburst, these are not expected to
be viable fueling mechanisms because the speed of the outflowing
stellar ejecta is far in excess of the escape velocity from the star
cluster \citep{norman88}. \citet{norman88} argue that the high
mass-loss rates and the low mass-loss speeds of the winds from evolved
AGB stars may offer a steady flow of material that will be retained
within the potential well of the star cluster. It can then cool and
(some fraction) can be accreted by the supermassive black hole. Thus,
the growth phase of the supermassive black hole (a powerful AGN) is
expected to occur only after the time that the mechanical feeedback
energy input from core collapse supernovae has subsided and the
progenitors of AGB stars have reached the end of their main sequence
lifetimes. This will occur at $\sim$50 Myr after the onset of the
starburst \citep{leitherer95}, and this age constraint is consistent
with the observed ages of the nuclear starbursts observed in a sample
of local AGN \citep{davies07}. Under the assumption of an
instantaneous burst, the ages of the DCOs shown in
Fig. \ref{fig:points} and listed in Table \ref{tab:massiveclumpprops}
are generally much lower (5--29 Myr) than 50 Myr. Thus, it is perhaps
not unexpected that under such a scenario the DCOs are currently not
capable of efficiently fuelling a supermassive black hole.

We also note that in the ``migrating--clump model'' advocated by
\citet{bournaud07,bournaud08} and \citet{elmegreen08b}, supermassive
nuclear black holes may form under the assumption that each of the
clumps gave rise to an intermediate mass black hole prior to its
migration toward the center where they can coalesce. Such a mechanism
may be able to explain the presence of seed black holes of substantial
mass in the centers of gas-rich galaxies dominated by massive,
turbulence-induced star-forming clumps. Both mechanisms described
above may therefore predict significant delay times between the onset
of a starburst and that of AGN.

\section{Summary and concluding remarks}

We have presented the results of the analysis of a sample of 30 
low-redshift ($z <0.3$) galaxies selected from the union of the GALEX
and SDSS surveys to have the high far-UV luminosities and small sizes
characteristic of the high-redshift Lyman Break Galaxies
\citep{heckman05,hoopes07,basu-zych07,overzier08}. These ``Lyman Break
Analogs'' (LBAs) offer the opportunity to investigate the physical
processes occurring in Lyman Break Galaxies (LBGs) in considerably
more detail than is possible at
high-redshift\footnote{\citet{heckman05} coined the term ``living
  fossils'' in relation to the Lyman break analogs sample then
  discovered. Indeed, an apt analogy in the field of evolutionary
  biology is presented by the Coelacanth. A modern-day analog of this
  species of fish, known from the fossil record and believed to be
  extinct since the end of the Cretaceous period, was discovered in
  1938. The find illustrates the use of analogs in the study of
  cosmology. Even though the local analogs will differ from their
  distant look-a-likes to some level of detail, and even though their
  local environment has probably changed drastically, they offer an
  entirely new way of investigating the fossil record alongside with
  other, more traditional lines of research.}. In this paper we
reported on the internal structures of the LBA's using HST optical and
UV images. We complemented these data with mid/far-IR photometry and
optical spectroscopy from the SDSS and the VLT.

We have shown that the morphologies of the LBAs are dominated by
massive compact clumps of young stars. {\it Our most unexpected and
  potentially important result is that we have found six galaxies in
  which the galaxy UV morphology is dominated by a single very compact
  object which is unresolved or only marginally resolved in the HST
  images.} These dominant compact objects (``DCOs'') have estimated
masses of few $\times 10^8$ to few $\times 10^9$ M$_{\odot}$ and radii
of-order 100 pc or less. In the other LBAs we instead find multiple
clumps per galaxy, the brightest of which have estimated masses of
typically a few $\times 10^8$ $M_{\odot}$ and radii of several hundred
pc.

We show that the galaxies containing the DCOs differ from the other
LBAs in many respects:\\

1. Their host galaxies tend to be more massive
than the other LBAs (median masses of $\sim 3 \times 10^{10}$
vs. $\sim 6 \times 10^9$ $M_{\odot}$ respectively).\\ 

2. They have systematically higher central concentrations
of light than the other LBAs (the DCO is typically located near the
center of a larger disk or envelope).\\ 

3. The DCOs are found in
galaxies with high mid-infrared luminosities: L$_{24\mu m} \sim
10^{10.3}$ to $\sim 10^{11.2}$ $L_{\odot}$, roughly an
order-of-magnitude more luminous than the other LBAs and comparable to
luminous infrared galaxies.\\ 

4. The galaxies with a DCO are more
strongly displaced from the locus of normal star forming galaxies in
the standard BPT diagnostic emission-line ratio diagram. This
displacement is in the sense of having larger \oiii/H$\beta$ and/or
larger \nii/H$\alpha$ ratios.\\

5. The DCOs tend to occur in galaxies
with higher gas densities and pressures as traced by the ionized gas
($P/k \sim 10^7$ cm$^{-3}$ K).\\ 

6. The DCOs are found in galaxies
in which the star-formation rate inferred from the dust-corrected
H$\alpha$ luminosity tends to be smaller than that derived from the
mid-IR luminosity and/or the extinction-corrected far-UV
luminosity. In the most extreme cases, this discrepancy reaches
roughly an order-of-magnitude.\\

We have shown that the DCOs are not Type 1 (unobscured) AGN. We find
no evidence for the presence of a very broad (several thousand km/s)
H$\alpha$ emission-line, with the most stringent constraints provided
by our VLT GIRAFFE/ARGUS IFU spectra. Instead, these data show
blue-asymmetric wings on the H$\alpha$ and \nii\ line profiles
indicative of a dusty outflow of ionized gas. While we can not
entirely exclude the possibility that an energetically significant
Type 2 (obscured) AGN is present in some or all of the DCOs, we regard
this possibility as unlikely. In particular, the DCOs are associated
with galaxies having unusually weak \sii\ and \oi\ emission-lines
compared to normal star forming galaxies. This is the opposite of what
is seen in starburst-AGN composite systems \citep[e.g.][]{kewley06}, but similar to
what is seen in low redshift IR-warm starbursts \citep{kewley01}.

The continuum of the DCOs is dominated by the light of young stars:
these are clearly strong highly compact starbursts. This is consistent
with both the high gas pressures and emission-line profiles: both are
commonly observed in starbursts and are the signposts of
supernova-driven galactic winds \citep[e.g.][]{lehnert96}. We have
considered two specific scenarios under which the starbursts
associated with the DCOs might be unusual. They may be evolved
short-duration starbursts in which the rate of photoionization is
relatively low because few O stars are present. They may also be
``leaky'' starbursts in which a significant fraction of the ionizing
radiation escapes (matter bounded conditions). The latter scenario
seems less likely, since it is difficult to envision circumstances in
which most of the ionizing radiation escapes the galaxy, but most of
the non-ionizing UV radiation is absorbed by dust and reprocessed in
the mid/far IR.

We have compared the properties of the DCOs and other massive clumps
to those of the nuclear star clusters typically seen in different
kinds of nearby galaxies. The DCOs are highly massive and dense
objects that do not appear to be related to single, young star
clusters which are significantly less massive ($\leq 10^7$
M$_{\odot}$) 
and smaller ($\sim$10 pc) than the structures we see in the
LBAs. However, the DCO structures are quite similar to predictions for
the merger-driven starbursts that are believed to be the progenitors
of the extra light or central cusps seen generically in lower-mass
($\leq M_*$) elliptical galaxies \citep{kormendy08,hopkins09}. This
would be consistent with our optical images of the LBAs, most of which
strongly suggest on-going mergers. However, at present it is
  unclear whether the DCO hosts could passively evolve to form a
  full-grown elliptical without additional merging.

More generally, the properties of the LBAs are quite consistent with
the idea that large Jeans instabilities in a gas-rich disk will lead
to the formation of highly massive star forming clumps. At high
  redshift, it is not yet clear what the main source of this gas is: it
  could be replenished continuously, or come in the form of discrete
  accretion events of small galaxies or clouds (or, most likely, a
  combination thereof). These processes have
  been proposed to explain the properties of high-$z$ star-forming
  galaxies \citep[e.g., see][and references
    therein]{overzier08,basu-zych09,elmegreen09,forster09,law07,law09}.
In any event, it is expected that the massive stellar clumps will 
aggregate at the center of a galaxy via dynamical friction to
ultimately form a bulge and supermassive black hole \citep[][and
  references therein]{elmegreen09}.  The LBAs offer us
the opportunity to study such processes in more detail, and compare
samples across different redshifts.

We speculated about why we do not see clear evidence for the formation
or rapid growth of a supermassive black hole in the DCOs (which appear
to be the ideal black hole nurseries). One intriguing possibility is that these
objects are simply too young for a black hole to be forming
yet. Following \citet{norman88} and \citet{davies07}, the growth of a
supermassive black hole inside a very massive compact star cluster may
not commence until the cluster is old enough for the core-collapse
supernovae phase to end (cluster age $>$ 50 Myr). Subsequent to this,
gas shed by evolved stars will be retained in the cluster's potential
well rather than being blown out. This gas can then cool, flow inward,
and be used to grow the black hole. If correct, we should be able to
identify a population of the descendants of the young DCOs in a phase
as powerful AGN. This scenario might also account for the rarity of
strong AGN in the LBG population \citep[e.g.][]{steidel02,ouchixx}.

Follow-up UV spectroscopy using the HST Cosmic Origins Spectrograph
has been allocated for selected targets in Cycle 17. The UV spectra
will allow us to study many of the features discussed in this paper
that are relevant to LBAs and to starbursts in general.   
These include the role of galactic outflows and low-luminosity AGN, 
the escape of ionizing radiation, the strength of
Ly$\alpha$\ emission, stellar abundances and the initial mass
function. 

%{\it Facilities:} \facility{HST (ACS)}, \facility{SSC (IRAC,MIPS)}.

\acknowledgments{We are very grateful to Frederic Bournaud, Rychard
  Bouwens, Jarle Brinchmann, Bruce Elmegreen, Guinevere Kauffmann,
  Lisa Kewley, Isa Oliveira, Francesco Shankar and the anonymous referee for useful
  suggestions and discussions. We thank Brent Groves for generating 
  the optically thin models referred to in Section 4. We thank Anne
  Pellerin for her help with the WFPC2 reductions. We thank the
  support staff at ESO Paranal for their assistance with the FLAMES
  observations.}

%\end{document}

\clearpage

\begin{deluxetable}{lrrcccccccllccc}
\tabletypesize\tiny
\tablecolumns{15}
\tablewidth{0pc}
\tablecaption{\label{tab:galprops}Properties of the Lyman break analogs sample.}
\tablehead{
\multicolumn{1}{c}{ID} &
\multicolumn{1}{c}{$\alpha_{J2000}$} &
\multicolumn{1}{c}{$\delta_{J2000}$} & 
\multicolumn{1}{c}{$z$} &
\multicolumn{1}{c}{log$M_*$}  & 
\multicolumn{1}{c}{$Z$$^f$}  & 
\multicolumn{3}{c}{Star Formation Rate} &
\multicolumn{1}{c}{log$L_{24}$} & 
\multicolumn{1}{c}{$R_e$$^a$} & 
\multicolumn{1}{c}{$C$$^b$} & 
\multicolumn{1}{c}{$\sigma_{v}$$^c$} &
\multicolumn{1}{c}{log$M_{dyn}$$^d$} & 
\multicolumn{1}{c}{$t_{dyn}$$^e$}\\
 & & & & \multicolumn{1}{c}{($M_\odot$)} & &
 \multicolumn{3}{c}{($M_\odot$ yr$^{-1}$)} & 
\multicolumn{1}{c}{($L_\odot$)} & \multicolumn{1}{c}{(kpc)} & &\multicolumn{1}{c}{(km s$^{-1}$)} &
 \multicolumn{1}{c}{($M_\odot$)} & \multicolumn{1}{c}{(Myr)}\\
 & & & & & & $H\alpha,0$ & $FUV,0$ & \ha$+$24 &  & & & & &}
\startdata
001009& 00:10:09.97 & --00:46:03.66 &     0.243 &      10.5&8.44&    4.6&   25.3&      4.8&        9.4&      3.06&      3.73&     91 &    10.3&     33\\
001054& 00:10:54.85 &     00:14:51.35 &     0.243&      -- &8.61&   14.4&    7.9&     26.9&       10.5&      4.24&      3.91&     157&   10.9&     27\\
004054& 00:40:54.33 & 15:34:09.66 &         0.283&      9.2&8.03&    7.4&    0.4&      13.9&       10.0&     0.95&      2.22&      65&    9.5&     14\\
005439& 00:54:39.80 & 15:54:46.93 &         0.236&     10.6&8.66&    4.6&    3.6&      15.4&       10.3&      1.73&      4.74&     120&   10.3&     14\\
005527& 00:55:27.46 & --00:21:48.71 &     0.167&        9.7&8.28&   22.7&   13.2&      55.4&       10.8&     0.77&      3.53&      124&    10.0&     6\\
015028& 01:50:28.41 & 13:08:58.40 &         0.147&     10.3&8.39&   19.4&   17.4&      50.7&       10.8&      1.83&      4.39&     103&   10.2&     17\\
020356& 02:03:56.91 & -08:07:58.51 &       0.189&       9.4&8.21&    11.1&   5.2&      14.7&       10.0&      1.61&      2.91&     77&   9.9&     21\\
021348& 02:13:48.54 & 12:59:51.46 &         0.219&     10.5&8.74&    6.7&   61.6&      35.1&       10.7&      1.53&      8.34&     86&   10.0&     18\\
032845& 03:28:45.99 & 01:11:50.85 &         0.142&      9.8&8.34&    6.7&    4.3&      8.7&        9.8&      1.82&      3.34&        83&   10.0&     22\\
035733& 03:57:34.00 & -05:37:19.70 &       0.204&      10.0&8.43&    9.6&   20.4&      12.7&       10.0&      1.09&      2.92&     86&   9.8&     12\\
040208& 04:02:08.87 & -05:06:42.06 &       0.139&       9.5&8.30&    2.0&    1.4&      2.5 &        9.0&      1.42&      3.92&          66&   9.7&     21\\
080232& 08:02:32.35 & 39:15:52.68 &         0.267&     10.7&8.65&   21.3&   75.7&      30.4&       10.5&      3.80&      4.87&    75 &   10.2&     50\\
080844& 08:08:44.27 & 39:48:52.36 &         0.091&      9.8&8.74&    3.7&   26.3&      16.1&       10.3&     0.88&      6.69&      131&   10.1&     7\\
082001& 08:20:01.72 & 50:50:39.16 &         0.217&      9.8&8.15&   20.3&    8.7&      40.0&       10.6&      1.52&      2.84&     91&   10.0&     17\\
082550& 08:25:50.95 & 41:17:10.30 &         0.156&      9.9&8.37&    5.5&    7.8&      7.0&        9.7&     1.56&      2.64&          94&   10.0&     17\\
083803& 08:38:03.73 & 44:59:00.28 &         0.143&      9.5&8.18&    4.5&    4.0&      6.2&         9.5&     0.92&      2.49&      75&   9.6&     12\\
092159& 09:21:59.39 & 45:09:12.38 &         0.235&     10.8&8.67&   25.8&   38.5&      55.1&       10.8&      1.80&      5.98&     135&   10.4&     13\\
092336& 09:23:36.46 & 54:48:39.25 &         0.222&      9.8&8.41&    9.3&    5.2&      9.9&       9.8&     0.48&      3.38&      101&   9.6&    5\\
092600& 09:26:00.41 & 44:27:36.13 &         0.181&      9.1&8.09&   13.2&    3.9&      17.0&       9.9&      1.09&      3.27&      95&   9.9&     11\\
093813& 09:38:13.50 & 54:28:25.09 &         0.102&      9.4&8.19&   13.0&    4.3&      19.8&       10.2&     0.92&      4.59&      98&   9.8&     9\\
101211& 10:12:11.18 & 63:25:03.70 &         0.246&      9.8&8.36&    4.3&   12.0&      6.2&       9.6&     0.89&      3.74&      79&   9.6&     11\\
102613& 10:26:13.97 & 48:44:58.94 &         0.160&      9.8&8.27&    7.0&    6.9&      9.9&       9.8&      1.99&      2.93&     70 &     9.6&     28\\
113303& 11:33:03.80 & 65:13:41.31 &         0.241&      9.1&8.02&    5.3&    3.9&      7.7&       9.6&     0.77&      3.30&     64 &   9.4&     12\\
124819& 12:48:19.75 & 66:21:42.68 &         0.260&      9.9&8.34&   17.8&   24.3&      18.4&       10.1&      1.90&      3.10&      71&   10.0&     24\\
135355& 13:53:55.90 & 66:48:00.59 &         0.198&      9.9&8.40&   17.1&   50.4&      19.4&       10.1&      3.57&      4.85&      86&   10.3&     41\\
143417& 14:34:17.16 & 02:07:42.58 &         0.180&     10.7&8.65&   25.4&   24.6&      20.0&       10.1&      4.60&      4.51&      99&   10.5&     46\\
210358& 21:03:58.75 & --07:28:02.45 &     0.137&       10.9&8.70&   41.3&    6.9&      108.3&       11.1&      2.70&      5.76&      173&   10.8&     15\\
214500& 21:45:00.26 & 01:11:57.58 &         0.204&      9.9&8.49&   13.6&   37.8&      16.4&       10.1&      1.16&      3.56&      78&   9.8&     15\\
231812& 23:18:13.00 & 00:41:26.10 &         0.252&     10.0&8.31&   30.8&   42.7&      63.1&       10.8&      2.54&      3.29&     82 &   10.1&     30\\
232539& 23:25:39.23 & 00:45:07.25 &         0.277&      9.2&8.18&    9.9&    9.5&      12.8&       9.9&     0.81&      2.93&      70&   9.5&     11\\
235347& 23:53:47.69 & 00:54:02.08 &         0.223&      9.5&8.11&    6.9&    2.1&      9.9&       9.8&      1.31&      2.27&      68&     9.7&     18\\
\enddata
\tablenotetext{a}{Optical half-light radius.}
\tablenotetext{b}{Optical concentration index.}
\tablenotetext{c}{Gas velocity derived from the \ha\ emission line width.}
\tablenotetext{d}{Dynamical mass $M_{dyn}=3.4R_e\sigma_{v}^2/G$.}
\tablenotetext{e}{Dynamical time $t_{dyn}=R_e/\sigma_{v}$.}
\tablenotetext{f}{Oxygen abundance in units of $12+log(O/H)$ estimated using the ``O3N2'' estimator from \citet{pettini04}.}
\end{deluxetable}

\begin{deluxetable}{lcccccccc}
\tabletypesize\tiny
\tablecolumns{8}
\tablewidth{0pc}
\tablecaption{\label{tab:lineprops}Emission line ratios and electron density.}
\tablehead{
\multicolumn{1}{c}{ID} & log(\oiiib/\hb) & log(\niib/\ha) & log(\siib/\ha)
& log(\oib/\ha) &log(\niib/\siib) &
\siib$\lambda$6713\AA/\siib$\lambda$6731\AA & $N_e$\\
& & & & & & & (cm$^{-3}$)}
\startdata
001009 &  0.29 &  -0.62 &  -- &  -1.09 &   -- &   -- &    --\\
001054 &  0.19 &  -0.19 &  -0.74 &  -1.98 &   0.55 &   1.45 &      40\\
004054 &  0.76 &  -1.43 &  -1.07 &  -1.60 &  -0.35 &   1.05 &     577\\
005439 &  -0.09&  -0.31 &  -0.62 &  -1.98 &   0.32 &   1.22 &     276\\
005527 &  0.56&  -0.84 &  -0.87 &  -1.56 &   0.03 &   1.10 &     485\\
015028 &  0.36&  -0.67 &  -0.70 &  -1.56 &   0.02 &   1.31 &     169\\
020356 &  0.53&  -1.08 &  -0.75 &  -1.60 &  -0.33 &   1.37 &     109\\
021348 &  -0.11&  -0.09 &  -0.63 &  -1.19 &   0.54 &   1.19 &     323\\
032845 &  0.41&  -0.82 &  -0.70 &  -1.63 &  -0.12 &   1.33 &     154\\
035733 &  0.21&  -0.74 &  -0.63 &  -1.54 &  -0.11 &   1.32 &     157\\
040208 &  0.41&  -0.93 &  -0.57 &  -1.48 &  -0.36 &   -- &     --\\
080232 &  -0.12&  -0.37 &  -0.54 &  -1.43 &   0.17 &   1.33 &     147\\
080844 &  -0.24&  -0.18 &  -0.82 &  -1.71 &   0.63 &   0.74 &    1991\\
082001 &  0.63&  -1.20 &  -0.79 &  -1.55 &  -0.41 &   1.27 &     219\\
082550 &  0.29&  -0.83 &  -0.55 &  -1.58 &  -0.27 &   1.45 &      36\\
083803 &  0.57&  -1.16 &  -0.79 &  -1.63 &  -0.37 &   1.31 &     173\\
092159 &  -0.13&  -0.30 &  -0.55 &  -1.21 &   0.25 &   1.18 &     346\\
092336 &  0.32&  -0.68 &  -0.66 &  -1.41 &  -0.02 &   1.25 &     246\\
092600 &  0.73&  -1.27 &  -1.03 &  -1.91 &  -0.24 &   1.31 &     169\\
093813 &  0.62&  -1.06 &  -0.87 &  -1.68 &  -0.18 &   1.33 &     154\\
101211 &  0.45&  -0.69 &  -0.70 &  -1.20 &   0.01 &   1.17 &     352\\
102613 &  0.40&  -1.03 &  -0.58 &  -1.42 &  -0.44 &   1.39 &      88\\
113303 &  0.74&  -1.46 &  -1.01 &  -1.98 &  -0.46 &   1.35 &     130\\
124819 &  0.38&  -0.85 &  -0.65 &  -1.59 &  -0.20 &   1.24 &     253\\
135355 &  0.24&  -0.79 &  -0.56 &  -1.48 &  -0.23 &   1.39 &      93\\
143417 &  -0.16&  -0.40 &  -0.60 &  -1.59 &   0.19 &   1.29 &     199\\
210358 &  -0.03&  -0.12 &  -0.74 &  -1.39 &   0.63 &   1.11 &     449\\
214500 &  0.10&  -0.64 &  -0.60 &  -1.48 &  -0.04 &   1.25 &     248\\
231812 &  0.44&  -0.87 &  -0.67 &  -1.42 &  -0.21 &   1.29 &     193\\
232539 &  0.62&  -1.10 &  -0.80 &  -1.57 &  -0.29 &   1.19 &     327\\
235347 &  0.69&  -1.25 &  -0.81 &  -1.55 &  -0.45 &   1.46 &      31\\
\enddata
\end{deluxetable}

\begin{deluxetable}{lrrcccccc}
\tabletypesize\small
\tablecolumns{9}
\tablewidth{0pc}
\tablecaption{\label{tab:clumpprops}Properties of the clumps.}
\tablehead{
\multicolumn{1}{c}{ID$^\dagger$} &
\multicolumn{1}{c}{\#} &
\multicolumn{1}{c}{Abs. Mag$^a$} & 
\multicolumn{1}{c}{Color$^b$} &
\multicolumn{1}{c}{Age$_{sb}$$^c$}  & 
\multicolumn{1}{c}{log$M_{*,sb}$$^c$} &
\multicolumn{1}{c}{Age$_{cst}$$^d$}  &
\multicolumn{1}{c}{log$M_{*,cst}$$^d$} &
\multicolumn{1}{c}{SFR$_{cst}$$^d$}\\
& & (mag) & (mag) & (Myr) & ($M_\odot$) &  (Myr) & ($M_\odot$) & ($M_\odot$ yr$^{-1}$)}
\startdata
210358 & 1 & -21.31 &  0.14 &      29 & 9.23 &     378 & 9.53 &  9.1 \\
021348 & 1 & -21.63 &  0.28 &      35 & 9.42 &     555 & 9.77 & 10.6 \\
092159 & 1 & -21.56 & -0.42 &      13 & 9.05 &      67 & 9.17 & 21.9 \\
080232 & 1 & -21.13 & -0.07 &      22 & 9.07 &     197 & 9.28 &  9.7 \\
080844$^\ddagger$ & 1 & -20.08 & -0.67 &       5 & 8.23 &       9 & 8.18 & 17.3\\
214500 & 1 & -19.91 &  0.34 &      38 & 8.76 &     647 & 9.13 &  2.1 \\
082550 & 1 & -18.93 &  0.76 &      63 & 8.49 &    1646 & 8.97 &  0.6 \\
082001 & 1 & -19.23 &  0.91 &      74 & 8.65 &    2044 & 9.12 &  0.7\\
124819 & 1 & -19.86 &  0.21 &      32 & 8.68 &     459 & 9.01 &  2.3\\
005439 & 1 & -19.85 &  0.06 &      26 & 8.61 &     295 & 8.88 &  2.6 \\
232539 & 1 & -19.79 &  0.22 &      32 & 8.65 &     467 & 8.99 &  2.1\\
231812 & 1 & -19.09 &  0.86 &      70 & 8.58 &    1909 & 9.06 &  0.6 \\
135355$^\ddagger$ & 1 & -19.42 &  0.09 &      21 & 8.42 &     222 & 8.69 &  2.2 \\
040208$^\ddagger$ & 1 & -18.06 &  0.43 &      53 & 8.16 &     560 & 8.36 &  0.5 \\
001009 & 1 & -19.73 &  0.01 &      24 & 8.54 &     256 & 8.79 &  2.5 \\
040208$^\ddagger$ & 2 & -17.90 &  0.49 &      60 & 8.13 &     626 & 8.32 &  0.4\\
102613$^\ddagger$ & 1 & -17.85 &  0.74 &      96 & 8.23 &     964 & 8.37 &  0.3\\
231812 & 2 & -19.28 &  0.14 &      29 & 8.41 &     372 & 8.72 &  1.4\\
005527$^\ddagger$ & 1 & -19.96 & -0.40 &       6 & 8.12 &      34 & 8.43 &  8.0\\
004054 & 2 & -18.89 &  0.60 &      53 & 8.43 &    1237 & 8.89 &  0.7\\
113303 & 1 & -19.34 & -0.15 &      19 & 8.31 &     155 & 8.50 &  2.1\\
040208$^\ddagger$ & 3 & -18.02 &  0.23 &      26 & 7.92 &     346 & 8.24 &  0.6\\
093813 & 1 & -19.15 & -0.60 &       5 & 7.67 &      35 & 8.03 &  3.1 \\
083803 & 1 & -18.27 & -0.24 &      17 & 7.85 &     121 & 8.00 &  0.9 \\
092336 & 1 & -20.38 & -0.55 &       6 & 8.15 &      42 & 8.57 &  8.9 \\
235347 & 1 & -18.34 &  0.37 &      40 & 8.14 &     695 & 8.52 &  0.5 \\
001054 & 1 & -20.49 & -0.52 &       6 & 8.19 &      48 & 8.65 &  9.4 \\
004054 & 1 & -18.43 &  0.55 &      50 & 8.24 &    1101 & 8.68 &  0.5 \\
143417 & 1 & -19.75 & -1.00 &       4 & 7.93 &       7 & 7.93 & 13.2 \\
020356 & 1 & -18.45 & -0.14 &      20 & 7.96 &     159 & 8.15 &  0.9 \\
015028 & 1 & -19.37 & -0.91 &       5 & 7.78 &       8 & 7.78 &  7.9 \\
231812 & 3 & -17.69 &  0.82 &      67 & 8.01 &    1796 & 8.49 &  0.2 \\
035733 & 1 & -19.03 & -0.44 &       6 & 7.59 &      64 & 8.14 &  2.2 \\
032845$^\ddagger$ & 1 & -17.77 & -0.47 &       6 & 7.26 &      25 & 7.48 &  1.3\\
032845$^\ddagger$ & 2 & -17.54 & -0.71 &       5 & 7.22 &       8 & 7.17 &  1.9 \\
135355$^\ddagger$ & 2 & -18.03 & -0.57 &       6 & 7.39 &      15 & 7.47 &  2.0\\
102613$^\ddagger$ & 2 & -17.71 & -0.19 &       6 & 7.18 &      83 & 7.76 &  0.7 \\
\enddata
\tablenotetext{$\dagger$}{ID number of the host galaxy of each clump.}
\tablenotetext{$\ddagger$}{Magnitudes and colors correspond to those measured in the filters F850LP and F330W--F850LP, instead of, F606W and F150LP--F606W, respectively.}
\tablenotetext{a}{Absolute (optical) clump magnitude from Fig. \ref{fig:cc},
  $M=m_{cor}-5\mathrm{log}\left(\frac{D_L}{10\mathrm{~pc}}\right)+2.5\mathrm{log}(1+z)$,
where $D_L$ is the luminosity distance. $m_{cor}$ is the apparent
magnitude corrected for dust using $m_{cor}=m-E(B-V)_*k^\prime(\lambda)$ and assuming
$E(B-V)_*=0.44E(B-V)_{gas}$ \citep{calzetti01}.}
\tablenotetext{b}{Dust-corrected clump color from Fig. \ref{fig:cc}.}
\tablenotetext{c}{Clump age and stellar mass under the assumption of the instantaneous star formation history shown in Fig. \ref{fig:cc} (red tracks).}
\tablenotetext{d}{Clump age, stellar mass and star formation rate under the assumption of the continuous star formation history shown in Fig. \ref{fig:cc} (blue tracks).}
\end{deluxetable}

\begin{deluxetable}{lcccccccccc}
\tabletypesize\small \tablecolumns{11} \tablewidth{0pc}
\tablecaption{\label{tab:massiveclumpprops}Properties of the massive,
  unresolved clumps (``DCOs'') discussed in Sect. \ref{sec:clumps}.}
\tablehead{ \multicolumn{1}{c}{ID$^\dagger$} &
  \multicolumn{1}{c}{$f_{UV}$$^a$} & \multicolumn{1}{c}{$f_{OPT}$$^b$}
  & \multicolumn{1}{c}{log$\langle M_{*,clump}\rangle$$^c$} &
  \multicolumn{1}{c}{$R_e$$^d$} &
  \multicolumn{1}{c}{log$\Sigma_M$$^e$} &
  \multicolumn{1}{c}{$\sigma_v$$^f$} &
  \multicolumn{1}{c}{$t_{dyn}$$^g$} &
  \multicolumn{1}{c}{$t_{fric}$$^h$} & log$\langle
  M_{*,gal}\rangle$$^i$ & $\langle M_{clump}\rangle/\langle
  M_{*,gal}\rangle$$^j$\\ &&& ($M_\odot$) & (pc) & ($M_\odot$
  kpc$^{-2}$) & (km s$^{-1}$) & (Myr) & (Myr) & ($M_\odot$) & }
\startdata 
210358 & 0.48 & 0.20 & 9.41 & 153 & 10.24 & 120 & 1 & 65 &10.9 & 0.04\\ 
021348 & 0.49 & 0.36 & 9.63 & 125 & 10.64 & 171 & 1 & 77& 10.3 & 0.23\\ 
092159 & 0.46 & 0.29 & 9.11 & 202 & 9.70 & 74 & 3 & 50& 10.6 & 0.03\\ 
080232 & 0.20 & 0.17 & 9.18 & 156 & 10.00 & 91 & 2 &153 & 10.5 & 0.04\\ 
080844 & 0.77 & 0.30 & 8.21 & 95 & 9.46 & 38 & 2 &56 & 10.0 & 0.02\\ 
005439 & 0.43 & 0.23 & 8.77 & 190 & 9.41 & 52 & 4 &70 & 10.5 & 0.02\\ 
\enddata 
\tablenotetext{$\dagger$}{ID number of the host galaxy of the DCO.}  
\tablenotetext{a,b}{Fractions of
  DCO-to-total light meaured in the UV ($f_{UV}$) and optical
  ($f_{OPT}$). A circular aperture of $0\farcs2$ radius was used for
  the DCO, and an aperture of 10\arcsec\ radius was used to estimate
  the total light. Companion galaxies were masked out and a large
  object-free annulus near the source was used for background
  subtraction. For object 080844 the UV/optical light fractions
  correspond to those measured in the filters F330W/F850LP, instead of
  F150LP/F606W. The total optical light for 080844 was measured within an
  aperture of 5\arcsec.}  
\tablenotetext{c}{Average of the single burst and continuous SFHs clump masses of Table \ref{tab:clumpprops}.}  
\tablenotetext{d}{Physical optical
  half-light radius measured within the $0\farcs4$ diameter aperture
  used in Sect. \ref{sec:sizes}.}  
\tablenotetext{e}{Stellar mass surface density $\Sigma_M=\langle M_{*,clump}\rangle/2\pi R_{clump}^2$.}  
\tablenotetext{f}{Clump velocity dispersion assuming
  spherical geometry $\sigma_{v,clump}=\sqrt{\frac{G\langle
      M_{*,clump}\rangle}{5R_{clump}}}$.}  
\tablenotetext{g}{Dynamical
  time $t_{dyn}=R_{clump}/\sigma_{v,clump}$}
\tablenotetext{h}{Dynamical friction time
  $t_{fric}=\left(\frac{2.64\times10^{11}}{\frac{1}{2}\mathrm{ln}(1+\Lambda^2)}\right)\left(\frac{R_{gal}}{2\mathrm{~kpc}}\right)^2
  \left(\frac{\sigma_{v,gal}}{250
    \mathrm{~km~s}^{-1}}\right)\left(\frac{10^6 M_\odot}{\langle
    M_{*,clump}\rangle}\right)$, where
  $\Lambda=\frac{R_{gal}\sigma_{v,gal}^2}{G\langle
    M_{*,clump}\rangle}$.}  \tablenotetext{i}{Galaxy stellar mass
  estimated from the average of the SED and dynamical masses.}
\tablenotetext{j}{Stellar mass fraction of the clumps.}
\end{deluxetable}

\end{document}